\documentclass[aps,prl,twocolumn,superscriptaddress]{revtex4-2}
\usepackage{amsmath,amssymb,mathtools}
\usepackage{bbm}
\usepackage{graphicx}
\usepackage{booktabs}
\usepackage{tikz}
\usetikzlibrary{arrows.meta,calc,positioning}
\usepackage{hyperref}
\usepackage{titletoc}
\usepackage{xcolor}

\begin{document}

\let\origaddcontentsline\addcontentsline
\renewcommand{\addcontentsline}[3]{}

\title{Extracting Boundary Conformal Data from Periodic Non-Hermitian Critical Chains}

\author{Yifan Liu}
\email{yifan@issp.u-tokyo.ac.jp}
\affiliation{Institute for Solid State Physics, University of Tokyo, Kashiwa, Chiba 277-8581, Japan}

\author{Haruki Shimizu}
\affiliation{Institute for Solid State Physics, University of Tokyo, Kashiwa, Chiba 277-8581, Japan}

\author{Dongchang Liu}

\affiliation{Mathematical Sciences Institute, The Australian National University, Canberra, ACT 2601, Australia}

\author{Kohei Kawabata}
\affiliation{Institute for Solid State Physics, University of Tokyo, Kashiwa, Chiba 277-8581, Japan}

\date{\today}

\begin{abstract}
Boundary conformal field theory (BCFT) contains universal data that are usually accessed microscopically by imposing spatial boundaries on the lattice. 
In non-Hermitian many-body systems, however, changing boundary conditions can qualitatively reorganize spectra and eigenstates, and their boundary criticality remains elusive.
Here, we introduce a periodic-chain spectroscopy to extract universal boundary quantities, such as the Affleck--Ludwig $g$-factor, directly from non-Hermitian bulk-critical quantum chains, 
avoiding the need to engineer microscopic open boundaries and circumventing subtle boundary effects in non-Hermitian systems. 
We illustrate our method with a $\mathcal{PT}$-symmetric Ising realization of the real nonunitary Yang--Lee CFT and reveal a universal negative excited-to-ground ratio, which has no counterpart in unitary critical theories and provides a microscopic signature of nonunitarity.
For the genuinely complex fixed points of the non-Hermitian five-state Potts chain, we extract intrinsically complex boundary coefficients, verify the exact Kramers--Wannier duality relation, and select the consistent analytic-continuation branch of the boundary states.
Our results establish a route to nonunitary BCFT universal data using only knowledge of the bulk critical system, opening a window into non-Hermitian boundary criticality.
\end{abstract}

\maketitle

\par\medskip\textit{Introduction.---}\ignorespaces
Boundaries and impurities can host universal critical behavior that is not determined by the bulk theory alone. Even when two systems share the same bulk critical point, different edges or localized perturbations can drive the boundary to distinct fixed points, leading to different finite-size spectra and responses.
Boundary conformal field theory (BCFT) characterizes these fixed points through universal boundary coefficients~\cite{CardyBoundary1989,AffleckLudwig1991,CardyBCFT2006}.
The best-known example is the Affleck--Ludwig $g$-factor, whose logarithm defines the boundary entropy and measures the effective degrees of freedom localized at the boundary~\cite{AffleckLudwig1991,FriedanKonechny2004}.
Beyond this single number, the full set of boundary coefficients determines the boundary partition function and constrains the excitation spectrum.
It therefore provides a universal fingerprint that identifies the realized boundary fixed point.

Microscopically, boundary conformal properties are commonly accessed by
imposing explicit open boundaries, through open-chain spectra or
boundary-sensitive partition-function, entanglement, and wave-function
observables~\cite{TangQMC2017,Stephan2010,AresRajabpourViti2020,Zou2022}. This reliance on imposed boundaries becomes problematic in
non-Hermitian critical systems.
Changing from periodic to open boundary conditions can qualitatively
reorganize spectra and eigenstates, as broadly exemplified by the
non-Hermitian skin effect~\cite{TonyLee2016,YaoWang2018,Kunst2018,ChingHuaLee2019,
YokomizoMurakami2019,KaiZhang2020,Okuma2020}.
Entanglement-based diagnostics
may also exhibit modified scaling in non-Hermitian open chains~\cite{PoYaoChang2020,ShimizuKawabata2025,VanderLinden2026}.
This makes it difficult to extract boundary coefficients while
maintaining direct control over the critical behavior established under
periodic boundary conditions.
This issue becomes increasingly relevant as the study of
unconventional criticality expands from exactly solvable
non-Hermitian chains~\cite{IkhlefJacobsenSaleur2012,FrahmSeel2014,
BazhanovKotousovKovalLukyanov2021,
RobertsonJacobsenSaleur2021,FrahmGehrmannKotousov2024}
and exceptional-point logarithmic CFTs~\cite{IoHuangHsieh2026}
to complex fixed points with intrinsically complex conformal data~\cite{Kaplan2009,Gorbenko2018,GorbenkoPotts2018,MaHe2019,
Haldar2023,Jacobsen2024,TangBulk2024,
ShimizuKawabata2025,VanderLinden2026}.
Although open-chain spectroscopy has identified candidate complex
conformal boundary conditions~\cite{TangBoundary2025}, it remains
unclear how to
extract universal boundary information without changing the critical system itself.

In this Letter, we address this problem and introduce a periodic-chain projected-partition-function spectroscopy.
Building on the
regularized-boundary-state picture relating short-range-entangled (SRE) states to conformal boundary states~\cite{DateJimboMiwaOkado1987,SaleurBauer1989,Foda2018,CardyBulkRG2017},
we formulate its non-Hermitian microscopic implementation.
Starting from a
periodic critical Hamiltonian and SRE preparation states, we
extract sector-resolved BCFT boundary coefficients $G_{a,i}$ from finite-size projections onto low-energy
periodic-chain eigenstates. 
For certain sectors,
these amplitudes
can be related to the generalized boundary entropy in
non-Hermitian systems. 
A crucial ingredient is a microscopically specified,
symmetry-compatible left-right pairing that defines the dual
preparation used in the projection.
For the protocols considered here, this pairing retains the nontrivial
signs and intrinsic complex phases of nonunitary boundary coefficients,
information inaccessible to ordinary positive-definite same-boundary
overlaps.
The construction therefore accesses a detailed boundary-state
fingerprint without engineering a separate open-chain Hamiltonian.

We demonstrate the method in two representative forms of
non-Hermitian criticality. For a $\mathcal{PT}$-symmetric Ising
realization of the Yang--Lee CFT
~\cite{YangLee1952,*LeeYang1952,Fisher:1978pf,CardyYangLee1985}, we recover the coefficient of the
stable conformal boundary condition together with a universal negative
excited-to-ground ratio. This sign has no counterpart in a unitary
theory with a positive-definite inner product and provides a direct
microscopic signature of nonunitarity.
We then study the non-Hermitian five-state Potts chain, whose periodic spectrum realizes 
complex-conjugate
fixed points~\cite{TangBulk2024}. 
There, we extract intrinsically complex boundary
coefficients and reproduce the duality relation $G_{\mathrm{free}}/G_{\mathrm{fixed}}=5$. By comparing these boundary coefficients with the
analytic continuation of conformal loop models~\cite{jacobsenConformalBoundaryLoop2008,dubailConformalTwoboundaryLoop2009,robertsonConformallyInvariantBoundary2019},
we identify the correct analytically continued fixed-boundary state, resolving the apparent
open-channel discrepancies involving fixed boundaries.
These results establish a consistent route from periodic non-Hermitian bulk-critical lattices to boundary
universal data, opening a new window into non-Hermitian boundary criticality.

\par\medskip\textit{Projected-partition-function spectroscopy.---}\ignorespaces
We now introduce our method for extracting boundary coefficients from non-Hermitian lattice models. 
In its explicit formulation (with field-theoretical validation
in Appendix~A), the inputs are a
periodic, bulk-critical non-Hermitian Hamiltonian $H$ and an
SRE state $\lvert \Phi\rangle$
that flows, in the
regularized-boundary-state sense, to a conformal boundary condition $a=a(\Phi)$ of the infrared CFT.
Different microscopic preparations may correspond to
the same fixed point $a(\Phi)$, and therefore share the same universal
projected coefficients while having different nonuniversal finite-size corrections.

Let $\lvert \Psi_{i,R}\rangle$ and $\langle \Psi_{i,L}\rvert$ be right and left low-energy eigenstates of the
periodic chain,
\begin{align}
  H\lvert \Psi_{i,R}\rangle
  =
  E_i\lvert \Psi_{i,R}\rangle ,
  \qquad
  \langle \Psi_{i,L}\rvert H
  =
  E_i\langle \Psi_{i,L}\rvert .
  \label{eq:left-right-eigenstates}
\end{align}
The low-energy spectrum is sorted according to the real part of the energy.
For each energy sector $i$, we define the paired projected amplitude~\footnote{If the energies are degenerate, one can alternatively define the projected amplitude to the subspace spanned by these degenerate eigenstates. An example is provided for the first excited state of the non-Hermitian five-state Potts model.}
\begin{align}
  Z_{\Phi,i}(L)
  \equiv
  \frac{
  \langle \widetilde{\Phi}| \Psi_{i,R}\rangle\langle \Psi_{i,L}| \Phi\rangle\,
  }{
  \langle \Psi_{i,L}| \Psi_{i,R}\rangle
  } ,
  \label{eq:paired-projected-amplitude}
\end{align}
where a compatible dual preparation $\langle \widetilde{\Phi}\rvert$ normalized as $\langle \widetilde{\Phi} | \Phi\rangle = 1$ is used due to non-Hermiticity. This dual state microscopically implements the nonunitary CFT bilinear pairing 
and can be determined via left-right
vector pairing using lattice symmetries, as detailed below.
Equation~\eqref{eq:paired-projected-amplitude} is the lattice analogue of a boundary cylinder amplitude projected onto the bulk conformal sector $i$. In
the scaling limit, the amplitude scales
\begin{align}
  Z_{\Phi,i}(L)
  =
  e^{-\alpha_\Phi  L}
  \left[
    G_{a(\Phi),i} + o(1)
  \right],
  \label{eq:Z-scaling}
\end{align}
where $\alpha_\Phi $ is a nonuniversal coefficient coming from the normalization, while $G_{a(\Phi),i}$ is a
universal boundary-state coefficient. Specifically, if the boundary state $|a\rangle$ has an expansion coefficient $B_a^{\,i}$ for the Ishibashi state $\lvert i\rangle\!\rangle$~\cite{Ishibashi1989}, then
\begin{align}
  G_{a,i} = \left(B_a^{\,i}\right)^2 .
  \label{eq:G-sector}
\end{align}
Thus our method extracts not only a single boundary entropy, but a set of sector-resolved coefficients
characterizing the boundary state.

The relation to the Affleck--Ludwig $g$-factor~\cite{AffleckLudwig1991} depends on the reference sector specified for the infrared
theory. 
In unitary theories, this is the identity sector, and Eq.~\eqref{eq:G-sector} gives
\begin{align}
G_{a,\mathbbm{1}} = g_a^2 .
\label{eq:identity-g}
\end{align}
In nonunitary theories, however, the leading cylinder amplitude is governed by the state with the lowest (real
part of) energy, which need not be the identity. We therefore denote this governing reference sector by
$\iota_g$, defining the boundary coefficient as
\begin{align}
G_a \equiv G_{a,\iota_g} .
\label{eq:g-sector}
\end{align}
The nonuniversal factor in Eq.~\eqref{eq:Z-scaling} can be removed by a finite-size fit,
\begin{align}
    -\log Z_{\Phi,\iota_g}(L) = \alpha_\Phi L - \log G_{a(\Phi)} + \frac{\beta_\Phi}{L} + \mathcal{O}(L^{-\omega}),
\end{align}
where $\beta_\Phi/L$ collects the leading finite-size correction from
the regularized boundary state, while
$\mathcal O(L^{-\omega})$ accounts for further irrelevant
perturbations (see Sec.~A of the Supplemental Material). Alternatively, for a more robust sector comparison that explicitly cancels the nonuniversal extensive term, one can use the ratio
\begin{align}
  R_{\Phi,i}(L)
  =
  \frac{Z_{\Phi,i}(L)}{Z_{\Phi,\iota_g}(L)}
  \longrightarrow
  \frac{G_{{a(\Phi)},i}}{G_{a(\Phi)}} .
  \label{eq:sector-ratio}
\end{align}
By preserving the sign or complex phase of the universal coefficient and remaining independent of the normalization of the preparation state, this ratio provides a
convenient diagnostic for non-Hermitian systems.

It remains to specify the dual preparation $\langle \widetilde{\Phi}\rvert$. The construction applies whenever
the microscopic model supplies a bilinear pairing of the left and right eigenstates compatible with the infrared CFT. We use two representative
classes. For an $\eta$-pseudo-Hermitian
realization~\cite{BenderBoettcher1998,Mostafazadeh2002PseudoHermiticity},
\begin{align}
  H^\dagger \eta = \eta H,
  \qquad
  \eta^\dagger = \eta ,
  \label{eq:eta-pseudo}
\end{align}
the natural choice is
\begin{align}
  \langle \widetilde{\Phi}\rvert
  =
  \frac{\langle \Phi\rvert \eta}
  {\langle \Phi\rvert \eta \lvert \Phi\rangle} .
  \label{eq:eta-dual-main}
\end{align}
This class describes real-spectrum non-Hermitian critical points, including the $\mathcal{PT}$-symmetric
Yang--Lee realization studied below~\footnote{When $\langle\Phi|\eta|\Phi\rangle=0$, as for the $Z$ preparation used
below, we use the unnormalized dual $\langle\Phi|\eta$ only in
normalization-independent sector ratios, where its overall factor
cancels.}.

The second class consists of genuinely complex fixed points that appear in conjugate pairs. If the lattice
Hamiltonian admits a symmetric bilinear form $\mathcal{S}_\sigma$ satisfying
\begin{align}
  H_\sigma^T \mathcal{S}_\sigma
  =
  \mathcal{S}_\sigma H_\sigma,
  \qquad
  \mathcal{S}_\sigma^T
  =
  \mathcal{S}_\sigma ,
  \label{eq:symmetric-pairing}
\end{align}
then the compatible dual preparation can be chosen as
\begin{align}
  \langle \widetilde{\Phi}^\sigma\rvert
  =
  \frac{
  \langle \mathcal{K}\Phi^\sigma\rvert \mathcal{S}_\sigma
  }{
  \langle \mathcal{K}\Phi^\sigma\rvert
  \mathcal{S}_\sigma
  \lvert \Phi^\sigma\rangle
  } ,
  \label{eq:complex-dual-main}
\end{align}
where $\mathcal{K}$ denotes complex conjugation. The non-Hermitian five-state Potts chain provides a simple
realization of this case, with $\mathcal{S}_\sigma=\mathbbm{1}$.

For the Hermitian case, the construction reduces to the standard bra-ket pairing by Hermitian conjugation
that is used widely in the literature~\cite{Stephan2010,Brockmann_2017}. 
As a consistency check, we provide transverse-field Ising and
three-state Potts benchmarks
in Sec.~B of the Supplemental Material~\cite{supplement}.

\par\medskip\textit{Yang--Lee signed boundary coefficients.---}\ignorespaces
We first validate the real branch of the construction by applying it to the non-Hermitian transverse-field Ising chain with periodic boundary conditions:
\begin{align}
  H_{\mathrm{YL}}
  =
  -\sum_{j=1}^{L}
  \left(
    \sigma_j^z\sigma_{j+1}^z
    +\lambda\sigma_j^x
    +ih\sigma_j^z
  \right).
  \label{eq:YL-hamiltonian}
\end{align}
This model possesses $\mathcal{PT}$ symmetry achieved by the combination of the parity operator $\mathcal{P}=\prod_j\sigma_j^x$ and the time-reversal operator $\mathcal{T} = \mathcal{K}$.
For each $\lambda>1$, there exists a critical field $h^{L}_c(\lambda)$ such that $\mathcal{PT}$ symmetry
is spontaneously broken and the universality class is identified with the $M(2,5)$ Yang--Lee
CFT~\cite{LeeYang1952,CardyYangLee1985,vonGehlen1991}. 
To avoid singularity at exceptional points,
finite-size projections $Z_{\Phi,i}(L)$ are then evaluated at the
infinite-size
value
$h=h_c^{(\infty)}(\lambda)$ extrapolated by finite-size exceptional points $h_c^L(\lambda)$. This also allows
us to stay in the $\mathcal{PT}$-unbroken phase where the model is $\eta$-pseudo-Hermitian with $\eta=\mathcal{P}$, so the prescription for the dual preparation in Eq.~\eqref{eq:eta-dual-main} applies.

The Yang--Lee CFT has two primaries $\mathbbm{1}$ and $\phi$, each associated with a Cardy boundary
state~\cite{CardyBoundary1989,DoreyRunkelTateoWatts2000,TakacsWatts2012,BajnokTompa2021, supplement}.
In this work, we only study the boundary coefficients associated
with the minimal-$g$ boundary condition $a=\mathbbm{1}$. The relevant universal
data are
\begin{align}
  G_{\mathbbm{1},\phi}
  =
  g_{\mathbbm{1}}^{\,2}=\sqrt{\frac{5-\sqrt5}{10}},
  \quad
  G_{\mathbbm{1},\mathbbm{1}}
  =-\sqrt{\frac{5+\sqrt5}{10}}.
  \label{eq:YL-ground-target}
\end{align}
The ratio
$G_{\mathbbm{1},\mathbbm{1}}/G_{\mathbbm{1},\phi}=-(1+\sqrt5)/2$ is a
nontrivial BCFT prediction,
which can be tested by
\begin{align}
  R_{\Phi,\mathbbm{1}}(L)
  =
  \frac{
    Z_{\Phi,\mathbbm{1}}(L)
  }{
    Z_{\Phi,\phi}(L)
  }
  \xrightarrow{L\to\infty}
  -\frac{1+\sqrt5}{2} .
  \label{eq:YL-ratio}
\end{align}
In particular, its negative sign cannot occur in
any unitary CFT, 
where reflection positivity ensures that each state-resolved
closed-channel amplitude is a nonnegative squared overlap
$|\langle i|a\rangle\!\rangle|^{2}$ between the boundary state and the
corresponding eigenstate, so all sector amplitudes share the same sign~\cite{yellowbook}. 
The
relative minus sign between the identity and $\phi$ sectors instead originates
from the negative-norm states of the indefinite inner product of the Yang--Lee
CFT~\cite{Fisher:1978pf}, and thus provides a direct and unambiguous signature of its nonunitarity.

\begin{figure}[t]
  \centering
  \includegraphics[width=\columnwidth]{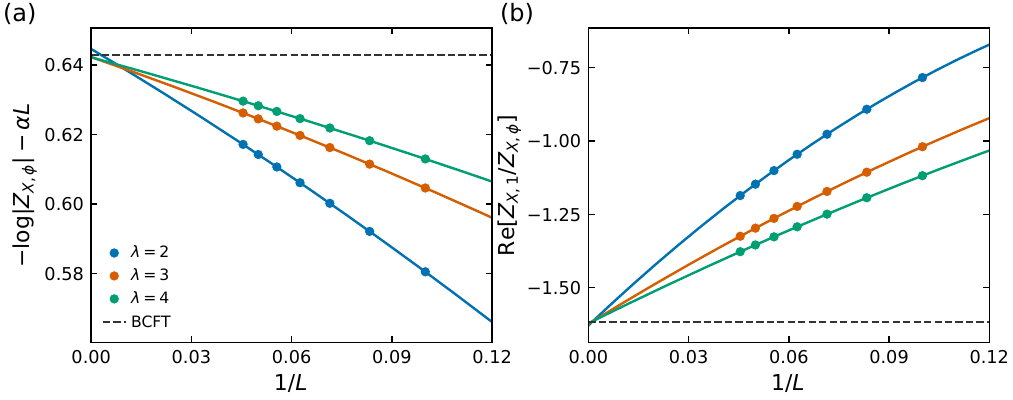}
  \caption{
  Yang--Lee boundary coefficients extracted from projected overlaps
  using the $\eta$-invariant
  preparation $\left|X\right\rangle$.
  The dashed lines are the BCFT predictions. The imaginary parts of both projected amplitudes are negligible.
  (a)~Extraction of
  $G_{\mathbbm{1}}=G_{\mathbbm{1},\phi}=g_{\mathbbm{1}}^2$ from the $\eta$-invariant
  $X$ preparation.
  (b)~The signed ratio
  $R_{X,\mathbbm{1}}=Z_{X,\mathbbm{1}}/Z_{X,\phi}$ cancels the extensive
  contribution and approaches $-(1+\sqrt5)/2$.
  }
  \label{fig:YL_boundary_coefficients}
\end{figure}

\begin{table}[t]
\caption{
\label{tab:YL-summary}
Yang--Lee benchmark at $\lambda=4$.
Here, $g_{\mathbbm{1}}^\mathrm{num}$ is the numerical extracted $g$-factor, and $\epsilon_\eta/\epsilon_0$ compares the relative imaginary
components of $R_\infty$ with and without the $\eta$ insertion.
The parentheses denote combined fit-window and
$h_c^{(\infty)}$ uncertainties.
}
\centering
\setlength{\tabcolsep}{4pt}
\renewcommand{\arraystretch}{1.05}
\begin{ruledtabular}
\begin{tabular}{llll}
prep. &
$g_{\mathbbm{1}}$ &
$R_\infty^{(\eta)}$ &
$\epsilon_\eta/\epsilon_0$ \\ \hline
$X$ &
$0.725339(51)$ &
$-1.62184(94)$ &
$10^{-10}/-$ \\
$Z$ &
-- &
$-1.61811(35)$ &
$10^{-10}/10^{-1}$ \\
Rand &
-- &
$-1.6218(18)$ &
$10^{-10}/10^{-1}$ \\
BCFT &
$0.725073$ &
$-1.618034$ &
--
\end{tabular}
\end{ruledtabular}
\end{table}

We use three product-state preparations: an $\eta$-invariant preparation $\left|X\right\rangle=\bigotimes_j(\left|\uparrow\right\rangle_j+\left|\downarrow\right\rangle_j)/\sqrt2$ for the extraction of $G_{\mathbbm{1},\phi}$; and two non-$\eta$-invariant preparations, $\left|Z\right\rangle=\bigotimes_j\left|\uparrow\right\rangle_j$ and a generic translation-invariant product state $\left|\mathrm{Rand}\right\rangle=\bigotimes_j\left|\theta,\phi\right\rangle_j$, for normalization-independent ratio diagnostics.
As shown in Fig.~\ref{fig:YL_boundary_coefficients} and summarized in Table~\ref{tab:YL-summary}, 
the $X$ preparation yields $g_{\mathbbm{1}}$ consistent with the BCFT value and accurately reproduces the negative-real projected ratio. 
Results for $\lambda=2,3$ and extrapolations are detailed in Sec.~C of the Supplemental Material~\cite{supplement}. 
Furthermore, for the non-$\eta$-invariant $Z$ and random preparations, the $\eta$-paired prescription suppresses the finite-size imaginary component to $\mathcal{O}(10^{-10})$, yielding a manifestly real estimator that cleanly approaches the expected ratio. This demonstrates the consistency of the $\eta$-paired protocol with the real universal boundary coefficients of the Yang--Lee BCFT.

The preparations considered here all converge to the same minimal-$g$ boundary condition ($a=\mathbbm{1}$),
giving identical universal boundary coefficients. This behavior suggests that the minimal-$g$ boundary
condition is the natural attractive fixed point under the renormalization group flow for a large class of
SRE preparations. Accessing other Yang--Lee boundary conditions with larger $g$-factor ($a=\phi$) requires
further microscopic control of the boundary preparation and is left for future work.

\par\medskip\textit{Non-Hermitian Potts complex boundary coefficients.---}\ignorespaces
To investigate a CFT characterized by
genuinely complex fixed points, we next apply our construction to the non-Hermitian five-state Potts chain~\cite{TangBulk2024,VanderLinden2026}. 
We study the system on the self-dual line at the complex fixed point $\lambda_c=0.079+0.060i$. 
The explicit Hamiltonian and numerical details are provided in Sec.~D of the Supplemental Material~\cite{supplement}.

The preparation states used here are tensor product states that preserve a subgroup of the full $S_5$-symmetry
group of the model. More explicitly, we use the family of preparations
\begin{align}
  \left|\Phi_k\right\rangle \equiv\bigotimes_{j=1}^L\frac{1}{\sqrt{k}}\sum_{n=0}^{k-1}\left|n\right\rangle
  \label{eq:potts-preparation-family}
\end{align}
to realize the standard $k$-color blob preparations of the Potts model~\cite{jacobsenConformalBoundaryLoop2008,dubailConformalTwoboundaryLoop2009,robertsonConformallyInvariantBoundary2019}.
With $\mathcal{S}_\sigma=\mathbbm{1}$ in this model, the dual preparation is given by $\langle
\widetilde{\Phi}_k|=\langle \Phi_k|$ according to Eq.~\eqref{eq:complex-dual-main}.

\begin{figure}[t]
  \centering
  \includegraphics[width=\columnwidth]{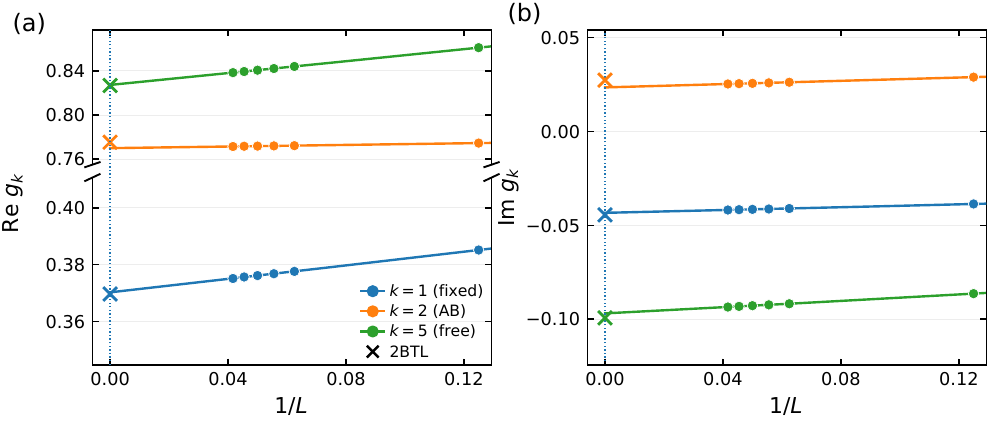}
  \caption{
  Finite-size extraction of complex Potts projected coefficients for representative preparations ($k=1,2,5$).
  We plot $g_{k,\mathrm{eff}}(L) \equiv
  \exp [(\log Z_{k,\mathbbm{1}}(L)+\alpha_kL)/2]$
  against $1/L$. The extrapolated
  intercept gives
  $g_k^\mathrm{num}$. The crosses at $1/L=0$ denote the corresponding
  boundary loop model values.
  }
  \label{fig:Potts_boundary_coefficients}
\end{figure}
\begin{table}[t]
  \caption{
  \label{tab:potts-G-family}
  Numerical and boundary-theory values of the complex Potts boundary
  coefficients $g_k=\sqrt{G_k}$ for the preparation family in
  Eq.~\eqref{eq:potts-preparation-family}.
  The parentheses denote the propagated fitting uncertainty.
  The relative deviation is
  $\delta_k=|g_k^{\rm num}-g_k^{\rm 2BTL}|/|g_k^{\rm 2BTL}|$.
  }
  \centering
  \begin{ruledtabular}
  \begin{tabular}{cccc}
    $k$ &
    $g_k^{\rm num}$ &
    $g_k^{\rm 2BTL}$ &
    $\delta_k$ \\ \hline
    $1\;(\mathrm{fixed})$ &
    $0.37018(3)-0.04363(3)\,i$ &
    $0.36967-0.04441\,i$ &
    $0.3\%$ \\ 
    $2$ &
    $0.7690(1)+0.0239(1)\,i$ &
    $0.7753+0.0273\,i$ &
    $0.9\%$ \\
    $3$ &
    $0.8881(5)-0.0258(5)\,i$ &
    $0.9473-0.0211\,i$ &
    $6\%$ \\
    $4$ &
    $0.8542(3)-0.0905(3)\,i$ &
    $0.8173-0.1245\,i$ &
    $6\%$ \\
    $5\;(\mathrm{free})$ &
    $0.82784(5)-0.09756(5)\,i$ &
    $0.82661-0.09930\,i$ &
    $0.3\%$
  \end{tabular}
  \end{ruledtabular}
\end{table}

The open-channel finite-size spectra of the complex Potts chain with conformal boundary
conditions were studied in Ref.~\cite{TangBoundary2025} and compared with the analytic
continuation of conformal loop models~\cite{jacobsenConformalBoundaryLoop2008,dubailConformalTwoboundaryLoop2009,robertsonConformallyInvariantBoundary2019}.
Here we perform the same analytic continuation in the complementary closed channel and compare the resulting boundary-theory coefficients with the numerically extracted complex boundary coefficients, 
as summarized in Table~\ref{tab:potts-G-family} (see Appendix~B for a detailed derivation).
Because these coefficients are sensitive to the branch of the analytically continued boundary
state, the fixed-boundary coefficient also selects the branch that removes the additional
higher open-channel characters reported in Ref.~\cite{TangBoundary2025}. In this sense, the
closed-channel boundary coefficient identifies the analytically continued fixed-boundary
state and clarifies the apparent open-channel discrepancies involving fixed boundaries.

Quantitatively, the fixed and free preparations show subpercent agreement with the two-boundary Temperley--Lieb (2BTL) predictions. They also satisfy the Kramers--Wannier duality relation
\cite{AffleckOshikawaSaleur1998,TangBoundary2025},
\begin{align}
  Z_{\text{free},\text{free}}
  =
  \sum_{n=0}^4 Z_{1,n}
  \Rightarrow
  G_{\text{free}}=5G_{\text{fixed}},
\end{align}
which is reproduced numerically with a $0.022\%$ relative complex deviation.
The intermediate preparations $k=2,3,4$, however, require more care. Although the
two-state mixed preparation $k=2$ agrees at the percent level in the ground-sector amplitude,
the first excited Potts multiplet analyzed in the Supplemental Material~\cite{supplement}
shows that this near agreement alone is not sufficient to establish the full analytically
continued 2BTL fingerprint. The residual
deviations for $k=2,3,4$ therefore point to a limitation of the naive analytic-continuation
assignment for intermediate blob boundaries, rather than merely to open-boundary finite-size
effects. A possible physical interpretation is suggested by the Hermitian four-state Potts
model, where perturbations of the fixed boundary condition are essentially irrelevant, so
breaking $S_4$ down to $S_3$ does not change the infrared boundary condition~\cite{liu2026b}.
If an analogous mechanism holds in the five-state complex Potts theory, the $k=3,4$
preparations may flow back to the free boundary condition in the scaling limit. The remaining independent
boundary conditions in this family would then obey the magnitude ordering
\begin{align}
  |g_{\text{free}}|>|g^{\rm num}_2|>|g_{\text{fixed}}|,
\end{align}
suggesting a possible generalized $g$-theorem for complex BCFT.

\par\medskip\textit{Summary and discussion.---}\ignorespaces
We have introduced a periodic-chain
spectroscopy for boundary universal data in non-Hermitian critical systems. The directly extracted observables are the boundary coefficients $G_{a,i}$;
the left boundary preparation is defined using a specified,
symmetry-compatible bilinear pairing rather than ordinary Hermitian
conjugation. This construction
recovers the boundary coefficient and a nontrivial negative ratio for the stable boundary condition in Yang--Lee BCFT. In the non-Hermitian
five-state Potts chain, it resolves intrinsically complex fixed and free boundary coefficients, and reproduces
the Kramers--Wannier relation. Notably, these closed-channel coefficients also identify the appropriate analytic-continuation branch of the boundary states, thereby clarifying the open-channel discrepancies that appear only at higher levels of the boundary spectrum. This suggests a broader diagnostic role for boundary coefficients in complex CFTs, where analytic continuation and branch choices often enter the construction of boundary data.

Our approach is complementary to the usual open-chain spectroscopy of boundary conformal towers:
it accesses boundary-state coefficients directly from periodic bulk-critical eigenstates, without
requiring a microscopic open-boundary Hamiltonian. The reliable results in the complex Potts case are also suggestive
from the perspective of boundary renormalization group flow.
Although no monotonic $g$-theorem is currently established for
complex BCFTs, the absolute values of the extracted $g$-factors are compatible with the ordering by effective boundary degrees of freedom. More broadly,
the same paired-overlap principle provides a route toward boundary conformal-data spectroscopy in
non-Hermitian systems, including optimized boundary preparations and bulk-to-boundary
operator-product coefficients, with
potential
applications to non-Hermitian quantum-quench problems.

\par\medskip\textit{Acknowledgments.---}\ignorespaces
We thank Yoshiki Fukusumi and Masaki Oshikawa for helpful discussions. 
We thank RIKEN iTHEMS, where this work was partially completed during the workshop on ``de Sitter Holography Meets Non-Hermitian Quantum Matter".
Y.~L. is supported by the Global Science Graduate Course (GSGC) program of the University of Tokyo.
D.~L. is supported by Australian Research Council Grant DP240100838.
K.~K. is supported by JSPS KAKENHI Grant No.~JP24H00945, No.~JP26H02015, No.~JP26K06970, and No.~JP26K17046, and JST FOREST Program Grant No.~JPMJFR256P. A part of the computation in this work was done using the facilities of the Supercomputer Center, the Institute for Solid State Physics, the University of Tokyo.

\bibliography{references}

@book{yellowbook,
  author = {Di Francesco, Philippe and Mathieu, Pierre and S{\'e}n{\'e}chal, David},
  title = {Conformal Field Theory},
  series = {Graduate Texts in Contemporary Physics},
  publisher = {Springer},
  address = {New York},
  year = {1997},
  doi = {10.1007/978-1-4612-2256-9}
}

@article{Fisher:1978pf,
  title = {{Yang--Lee} Edge Singularity and ${\ensuremath{\phi}}^{3}$ Field Theory},
  author = {Fisher, Michael E.},
  journal = {Physical Review Letters},
  volume = {40},
  number = {25},
  pages = {1610--1613},
  year = {1978},
  doi = {10.1103/PhysRevLett.40.1610}
}

@article{YaoWang2018,
  author = {Yao, Shunyu and Wang, Zhong},
  title = {Edge States and Topological Invariants of Non-{Hermitian} Systems},
  journal = {Physical Review Letters},
  volume = {121},
  number = {8},
  pages = {086803},
  year = {2018},
  doi = {10.1103/PhysRevLett.121.086803},
  eprint = {1803.01876},
  archivePrefix = {arXiv},
  primaryClass = {cond-mat.mes-hall}
}

@article{dubailConformalTwoboundaryLoop2009,
  title = {Conformal Two-Boundary Loop Model on the Annulus},
  author = {Dubail, J{\'e}r{\^o}me and Jacobsen, Jesper Lykke and Saleur, Hubert},
  year = {2009},
  journal = {Nuclear Physics B},
  volume = {813},
  number = {3},
  pages = {430--459},
  doi = {10.1016/j.nuclphysb.2008.12.023},
  eprint = {0812.2746},
  archivePrefix = {arXiv},
  primaryClass = {math-ph}
}

@article{jacobsenConformalBoundaryLoop2008,
  title = {Conformal Boundary Loop Models},
  author = {Jacobsen, Jesper Lykke and Saleur, Hubert},
  year = {2008},
  journal = {Nuclear Physics B},
  volume = {788},
  number = {3},
  pages = {137--166},
  doi = {10.1016/j.nuclphysb.2007.06.029},
  eprint = {math-ph/0611078},
  archivePrefix = {arXiv}
}

@article{robertsonConformallyInvariantBoundary2019,
  title = {Conformally Invariant Boundary Conditions in the Antiferromagnetic {Potts} Model and the {SL}(2, {{$\mathbb{R}$}})/{U}(1) Sigma Model},
  author = {Robertson, Niall F. and Jacobsen, Jesper Lykke and Saleur, Hubert},
  year = {2019},
  journal = {Journal of High Energy Physics},
  volume = {2019},
  number = {10},
  pages = {254},
  doi = {10.1007/JHEP10(2019)254},
  eprint = {1906.07565},
  archivePrefix = {arXiv},
  primaryClass = {cond-mat.stat-mech}
}

@article{liu2026b,
  title = {Boundary critical phenomena in the quantum {Ashkin--Teller} model},
  author = {Liu, Yifan and Chepiga, Natalia and Fukusumi, Yoshiki and Oshikawa, Masaki},
  year = {2026},
  journal = {SciPost Physics},
  volume = {20},
  number = {5},
  pages = {130},
  doi = {10.21468/SciPostPhys.20.5.130},
  eprint = {2601.16951},
  archivePrefix = {arXiv},
  primaryClass = {cond-mat.str-el}
}

@incollection{CardyBCFT2006,
author = {Cardy, John},
title = {Boundary Conformal Field Theory},
booktitle = {Encyclopedia of Mathematical Physics},
publisher = {Elsevier},
year = {2006},
doi = {10.1016/B0-12-512666-2/00398-9},
eprint = {hep-th/0411189},
archivePrefix = {arXiv}
}

@article{TonyLee2016,
  title = {Anomalous Edge State in a Non-{Hermitian} Lattice},
  author = {Lee, Tony E.},
  journal = {Physical Review Letters},
  volume = {116},
  number = {13},
  pages = {133903},
  year = {2016},
  doi = {10.1103/PhysRevLett.116.133903},
  eprint = {1603.05312},
  archivePrefix = {arXiv},
  primaryClass = {quant-ph}
}

@article{Kunst2018,
  author = {Kunst, Flore K. and Edvardsson, Elisabet and Budich, Jan Carl and Bergholtz, Emil J.},
  title = {Biorthogonal Bulk-Boundary Correspondence in Non-{Hermitian} Systems},
  journal = {Physical Review Letters},
  volume = {121},
  pages = {026808},
  year = {2018},
  doi = {10.1103/PhysRevLett.121.026808},
  eprint = {1805.06492},
  archivePrefix = {arXiv},
  primaryClass = {cond-mat.mes-hall}
}

@article{ChingHuaLee2019,
  title = {Anatomy of skin modes and topology in non-{Hermitian} systems},
  author = {Lee, Ching Hua and Thomale, Ronny},
  journal = {Physical Review B},
  volume = {99},
  number = {20},
  pages = {201103(R)},
  year = {2019},
  doi = {10.1103/PhysRevB.99.201103},
  eprint = {1809.02125},
  archivePrefix = {arXiv},
  primaryClass = {cond-mat.other}
}

@article{YokomizoMurakami2019,
  author = {Yokomizo, Kazuki and Murakami, Shuichi},
  title = {Non-{Bloch} Band Theory of Non-{Hermitian} Systems},
  journal = {Physical Review Letters},
  volume = {123},
  number = {6},
  pages = {066404},
  year = {2019},
  doi = {10.1103/PhysRevLett.123.066404},
  eprint = {1902.10958},
  archivePrefix = {arXiv},
  primaryClass = {cond-mat.mes-hall}
}

@article{KaiZhang2020,
  title = {Correspondence between Winding Numbers and Skin Modes in Non-{Hermitian} Systems},
  author = {Zhang, Kai and Yang, Zhesen and Fang, Chen},
  journal = {Physical Review Letters},
  volume = {125},
  number = {12},
  pages = {126402},
  year = {2020},
  doi = {10.1103/PhysRevLett.125.126402},
  eprint = {1910.01131},
  archivePrefix = {arXiv},
  primaryClass = {cond-mat.mes-hall}
}

@article{Okuma2020,
  author = {Okuma, Nobuyuki and Kawabata, Kohei and Shiozaki, Ken and Sato, Masatoshi},
  title = {Topological Origin of Non-{Hermitian} Skin Effects},
  journal = {Physical Review Letters},
  volume = {124},
  number = {8},
  pages = {086801},
  year = {2020},
  doi = {10.1103/PhysRevLett.124.086801},
  eprint = {1910.02878},
  archivePrefix = {arXiv},
  primaryClass = {cond-mat.mes-hall}
}

@article{BenderBoettcher1998,
  author = {Bender, Carl M. and Boettcher, Stefan},
  title = {Real Spectra in Non-{Hermitian} {Hamiltonians} Having $\mathcal{PT}$ Symmetry},
  journal = {Physical Review Letters},
  volume = {80},
  number = {24},
  pages = {5243--5246},
  year = {1998},
  doi = {10.1103/PhysRevLett.80.5243},
  eprint = {physics/9712001},
  archivePrefix = {arXiv}
}

@article{YangLee1952,
  title = {Statistical Theory of Equations of State and Phase Transitions. {I}. Theory of Condensation},
  author = {Yang, C. N. and Lee, T. D.},
  journal = {Physical Review},
  volume = {87},
  number = {3},
  pages = {404--409},
  year = {1952},
  doi = {10.1103/PhysRev.87.404}
}

@article{LeeYang1952,
  author = {Lee, T. D. and Yang, C. N.},
  title = {Statistical Theory of Equations of State and Phase Transitions. {II}. Lattice Gas and {Ising} Model},
  journal = {Physical Review},
  volume = {87},
  number = {3},
  pages = {410--419},
  year = {1952},
  doi = {10.1103/PhysRev.87.410}
}

@article{CardyYangLee1985,
  author = {Cardy, John L.},
  title = {Conformal Invariance and the {Yang--Lee} Edge Singularity in Two Dimensions},
  journal = {Physical Review Letters},
  volume = {54},
  number = {13},
  pages = {1354--1356},
  year = {1985},
  doi = {10.1103/PhysRevLett.54.1354}
}

@article{vonGehlen1991,
  author = {von Gehlen, G.},
  title = {Critical and Off-Critical Conformal Analysis of the {Ising} Quantum Chain in an Imaginary Field},
  journal = {Journal of Physics A: Mathematical and General},
  volume = {24},
  number = {22},
  pages = {5371--5399},
  year = {1991},
  doi = {10.1088/0305-4470/24/22/021}
}

@article{Kaplan2009,
  author = {Kaplan, David B. and Lee, Jong-Wan and Son, Dam T. and Stephanov, Mikhail A.},
  title = {Conformality Lost},
  journal = {Physical Review D},
  volume = {80},
  number = {12},
  pages = {125005},
  year = {2009},
  doi = {10.1103/PhysRevD.80.125005},
  eprint = {0905.4752},
  archivePrefix = {arXiv},
  primaryClass = {hep-th}
}

@article{Gorbenko2018,
  author = {Gorbenko, Victor and Rychkov, Slava and Zan, Bernardo},
  title = {Walking, weak first-order transitions, and complex {CFTs}},
  journal = {Journal of High Energy Physics},
  volume = {2018},
  number = {10},
  pages = {108},
  year = {2018},
  doi = {10.1007/JHEP10(2018)108},
  eprint = {1807.11512},
  archivePrefix = {arXiv},
  primaryClass = {hep-th}
}

@article{GorbenkoPotts2018,
  author = {Gorbenko, Victor and Rychkov, Slava and Zan, Bernardo},
  title = {Walking, Weak first-order transitions, and Complex {CFTs} {II}. Two-dimensional {Potts} model at $Q>4$},
  journal = {SciPost Physics},
  volume = {5},
  number = {5},
  pages = {050},
  year = {2018},
  doi = {10.21468/SciPostPhys.5.5.050},
  eprint = {1808.04380},
  archivePrefix = {arXiv},
  primaryClass = {hep-th}
}

@article{MaHe2019,
  author = {Ma, Han and He, Yin-Chen},
  title = {Shadow of complex fixed point: Approximate conformality of $Q > 4$ {Potts} model},
  journal = {Physical Review B},
  volume = {99},
  number = {19},
  pages = {195130},
  year = {2019},
  doi = {10.1103/PhysRevB.99.195130},
  eprint = {1811.11189},
  archivePrefix = {arXiv},
  primaryClass = {cond-mat.str-el}
}

@article{CardyBoundary1989,
  author = {Cardy, John L.},
  title = {Boundary Conditions, Fusion Rules and the {Verlinde} Formula},
  journal = {Nuclear Physics B},
  volume = {324},
  number = {3},
  pages = {581--596},
  year = {1989},
  doi = {10.1016/0550-3213(89)90521-X}
}

@article{Ishibashi1989,
  author = {Ishibashi, Nobuyuki},
  title = {The Boundary and Crosscap States in Conformal Field Theories},
  journal = {Modern Physics Letters A},
  volume = {4},
  number = {3},
  pages = {251--264},
  year = {1989},
  doi = {10.1142/S0217732389000320}
}

@article{AffleckLudwig1991,
  author = {Affleck, Ian and Ludwig, Andreas W. W.},
  title = {Universal Noninteger ``Ground-State Degeneracy'' in Critical Quantum Systems},
  journal = {Physical Review Letters},
  volume = {67},
  number = {2},
  pages = {161--164},
  year = {1991},
  doi = {10.1103/PhysRevLett.67.161}
}

@article{FriedanKonechny2004,
  author = {Friedan, Daniel and Konechny, Anatoly},
  title = {Boundary Entropy of One-Dimensional Quantum Systems at Low Temperature},
  journal = {Physical Review Letters},
  volume = {93},
  number = {3},
  pages = {030402},
  year = {2004},
  doi = {10.1103/PhysRevLett.93.030402},
  eprint = {hep-th/0312197},
  archivePrefix = {arXiv}
}

@article{Stephan2010,
  author = {St{\'e}phan, Jean-Marie and Misguich, Gr{\'e}goire and Alet, Fabien},
  title = {Geometric entanglement and {Affleck--Ludwig} boundary entropies in critical {XXZ} and {Ising} chains},
  journal = {Physical Review B},
  volume = {82},
  number = {18},
  pages = {180406(R)},
  year = {2010},
  doi = {10.1103/PhysRevB.82.180406},
  eprint = {1007.4161},
  archivePrefix = {arXiv},
  primaryClass = {cond-mat.stat-mech}
}

@article{Brockmann_2017,
  author = {Brockmann, Michael and St{\'e}phan, Jean-Marie},
  title = {Universal terms in the overlap of the ground state of the spin-1/2 {XXZ} chain with the {N{\'e}el} state},
  journal = {Journal of Physics A: Mathematical and Theoretical},
  volume = {50},
  number = {35},
  pages = {354001},
  year = {2017},
  doi = {10.1088/1751-8121/aa809c},
  eprint = {1705.08505},
  archivePrefix = {arXiv},
  primaryClass = {cond-mat.stat-mech}
}

@misc{IoHuangHsieh2026,
  author = {Io, Iao-Fai and Huang, Fu-Hsiang and Hsieh, Chang-Tse},
  title = {Non-{Hermitian} free-fermion critical systems and logarithmic conformal field theory},
  year = {2026},
  eprint = {2602.02649},
  archivePrefix = {arXiv},
  primaryClass = {cond-mat.str-el}
}

@article{TangQMC2017,
  author = {Tang, Wei and Chen, Lei and Li, Wei and Xie, X. C. and Tu, Hong-Hao and Wang, Lei},
  title = {Universal Boundary Entropies in Conformal Field Theory: A Quantum {Monte Carlo} Study},
  journal = {Physical Review B},
  volume = {96},
  number = {11},
  pages = {115136},
  year = {2017},
  doi = {10.1103/PhysRevB.96.115136},
  eprint = {1708.04022},
  archivePrefix = {arXiv},
  primaryClass = {cond-mat.str-el}
}

@article{DoreyRunkelTateoWatts2000,
  author = {Dorey, Patrick and Runkel, Ingo and Tateo, Roberto and Watts, G{\'e}rard M. T.},
  title = {$g$-Function Flow in Perturbed Boundary Conformal Field Theories},
  journal = {Nuclear Physics B},
  volume = {578},
  number = {1--2},
  pages = {85--122},
  year = {2000},
  doi = {10.1016/S0550-3213(99)00772-5},
  eprint = {hep-th/9909216},
  archivePrefix = {arXiv}
}

@article{TakacsWatts2012,
  author = {Tak{\'a}cs, G{\'a}bor and Watts, G{\'e}rard M. T.},
  title = {Excited-State $g$-Functions from the Truncated Conformal Space},
  journal = {Journal of High Energy Physics},
  volume = {2012},
  number = {2},
  pages = {082},
  year = {2012},
  doi = {10.1007/JHEP02(2012)082},
  eprint = {1112.2906},
  archivePrefix = {arXiv},
  primaryClass = {hep-th}
}

@article{AffleckOshikawaSaleur1998,
  author = {Affleck, Ian and Oshikawa, Masaki and Saleur, Hubert},
  title = {Boundary Critical Phenomena in the Three-State {Potts} Model},
  journal = {Journal of Physics A: Mathematical and General},
  volume = {31},
  number = {28},
  pages = {5827--5842},
  year = {1998},
  doi = {10.1088/0305-4470/31/28/003},
  eprint = {cond-mat/9804117},
  archivePrefix = {arXiv}
}

@article{Mostafazadeh2002PseudoHermiticity,
  author = {Mostafazadeh, Ali},
  title = {Pseudo-{Hermiticity} versus 
 {PT} symmetry: The necessary condition for the reality of the spectrum of a non-{Hermitian} {Hamiltonian}},
  journal = {Journal of Mathematical Physics},
  volume = {43},
  number = {1},
  pages = {205--214},
  year = {2002},
  doi = {10.1063/1.1418246},
  eprint = {math-ph/0107001},
  archivePrefix = {arXiv}
}

@article{Haldar2023,
  title = {Hidden Critical Points in the Two-Dimensional $\mathrm{O}(n>2)$ Model: Exact Numerical Study of a Complex Conformal Field Theory},
  author = {Haldar, Arijit and Tavakol, Omid and Ma, Han and Scaffidi, Thomas},
  journal = {Physical Review Letters},
  volume = {131},
  number = {13},
  pages = {131601},
  year = {2023},
  doi = {10.1103/PhysRevLett.131.131601},
  eprint = {2303.02171},
  archivePrefix = {arXiv},
  primaryClass = {cond-mat.stat-mech}
}

@article{Jacobsen2024,
  title = {Lattice Realization of Complex Conformal Field Theories: Two-Dimensional {Potts} Model with $Q> 4$ States},
  author = {Jacobsen, Jesper Lykke and Wiese, Kay J\"org},
  journal = {Physical Review Letters},
  volume = {133},
  number = {7},
  pages = {077101},
  year = {2024},
  doi = {10.1103/PhysRevLett.133.077101},
  eprint = {2402.10732},
  archivePrefix = {arXiv},
  primaryClass = {hep-th}
}

@article{TangBulk2024,
  author = {Tang, Yin and Ma, Han and Tang, Qicheng and He, Yin-Chen and Zhu, W.},
  title = {Reclaiming the Lost Conformality in a Non-{Hermitian} Quantum 5-State {Potts} Model},
  journal = {Physical Review Letters},
  volume = {133},
  number = {7},
  pages = {076504},
  year = {2024},
  doi = {10.1103/PhysRevLett.133.076504},
  eprint = {2403.00852},
  archivePrefix = {arXiv},
  primaryClass = {cond-mat.stat-mech}
}

@article{ShimizuKawabata2025,
  author = {Shimizu, Haruki and Kawabata, Kohei},
  title = {Complex Entanglement Entropy for Complex Conformal Field Theory},
  journal = {Physical Review B},
  volume = {112},
  number = {8},
  pages = {085112},
  year = {2025},
  doi = {10.1103/n578-ljd5},
  eprint = {2502.02001},
  archivePrefix = {arXiv},
  primaryClass = {cond-mat.stat-mech}
}

@article{TangBoundary2025,
  author = {Tang, Yin and Liu, Qianyu and Tang, Qicheng and Zhu, W.},
  title = {Boundary criticality of complex conformal field theory: A case study in the non-{Hermitian} 5-state {Potts} model},
  journal = {SciPost Physics},
  volume = {19},
  number = {6},
  pages = {164},
  year = {2025},
  doi = {10.21468/SciPostPhys.19.6.164},
  eprint = {2512.07625},
  archivePrefix = {arXiv},
  primaryClass = {cond-mat.stat-mech}
}

@article{VanderLinden2026,
  author = {Vander Linden, Vic and De Vos, Boris and Vervoort, Kevin and Verstraete, Frank and Ueda, Atsushi},
  title = {Spiral renormalization group flow and universal entanglement spectrum of the non-{Hermitian} five-state {Potts} model},
  journal = {Physical Review B},
  volume = {113},
  number = {20},
  pages = {205106},
  year = {2026},
  doi = {10.1103/l1cp-6gzr},
  eprint = {2507.14732},
  archivePrefix = {arXiv},
  primaryClass = {cond-mat.str-el}
}

@article{chepigaCriticalPropertiesQuantum2022,
  title = {Critical Properties of Quantum Three- and Four-State {Potts} Models with Boundaries Polarized along the Transverse Field},
  author = {Chepiga, Natalia},
  year = {2022},
  journal = {SciPost Physics Core},
  volume = {5},
  number = {2},
  eprint = {2107.08899},
  primaryClass = {cond-mat.str-el},
  pages = {031},
  doi = {10.21468/SciPostPhysCore.5.2.031},
  archivePrefix = {arXiv}
}

@article{BajnokTompa2021,
  author = {Bajnok, Zoltan and Tompa, Tamas Lajos},
  title = {{TCSA} and the finite volume boundary state},
  journal = {Nuclear Physics B},
  volume = {964},
  pages = {115330},
  year = {2021},
  doi = {10.1016/j.nuclphysb.2021.115330},
  eprint = {2008.01979},
  archivePrefix = {arXiv},
  primaryClass = {hep-th}
}

@article{Zou2022,
  author = {Zou, Yijian},
  title = {Universal Information of Critical Quantum Spin Chains from Wavefunction Overlap},
  journal = {Physical Review B},
  volume = {105},
  number = {16},
  pages = {165420},
  year = {2022},
  doi = {10.1103/PhysRevB.105.165420},
  eprint = {2104.00103},
  archivePrefix = {arXiv},
  primaryClass = {cond-mat.str-el}
}

@article{PoYaoChang2020,
  title = {Entanglement spectrum and entropy in topological non-{Hermitian} systems and nonunitary conformal field theory},
  author = {Chang, Po-Yao and You, Jhih-Shih and Wen, Xueda and Ryu, Shinsei},
  journal = {Physical Review Research},
  volume = {2},
  number = {3},
  pages = {033069},
  year = {2020},
  doi = {10.1103/PhysRevResearch.2.033069},
  eprint = {1909.01346},
  archivePrefix = {arXiv},
  primaryClass = {cond-mat.str-el}
}

@article{FSCYifan,
  title = {Finite-size corrections to the energy spectra of gapless one-dimensional systems in the presence of boundaries},
  author = {Liu, Yifan and Shimizu, Haruki and Ueda, Atsushi and Oshikawa, Masaki},
  journal = {SciPost Physics},
  volume = {17},
  number = {4},
  pages = {099},
  year = {2024},
  doi = {10.21468/SciPostPhys.17.4.099},
  eprint = {2405.06891},
  archivePrefix = {arXiv},
  primaryClass = {cond-mat.str-el}
}

@article{CalabreseCardy2006,
  author = {Calabrese, Pasquale and Cardy, John},
  title = {Time Dependence of Correlation Functions Following a Quantum Quench},
  journal = {Physical Review Letters},
  volume = {96},
  number = {13},
  pages = {136801},
  year = {2006},
  doi = {10.1103/PhysRevLett.96.136801},
  eprint = {cond-mat/0601225},
  archivePrefix = {arXiv}
}

@article{BelavinPolyakovZamolodchikov1984,
  author = {Belavin, A. A. and Polyakov, A. M. and Zamolodchikov, A. B.},
  title = {Infinite Conformal Symmetry in Two-Dimensional Quantum Field Theory},
  journal = {Nuclear Physics B},
  volume = {241},
  number = {2},
  pages = {333--380},
  year = {1984},
  doi = {10.1016/0550-3213(84)90052-X}
}

@misc{FukusumiKawamoto2026,
  author = {Fukusumi, Yoshiki and Kawamoto, Taishi},
  title = {Generalizing quantum dimensions: Symmetry-based classification of local {pseudo-Hermitian} systems and the corresponding domain walls},
  year = {2025},
  eprint = {2511.11059},
  archivePrefix = {arXiv},
  primaryClass = {hep-th}
}

@article{xu2022,
  title = {{2D} {Ising} Field Theory in a magnetic field: the {Yang--Lee} singularity},
  author = {Xu, Hao-Lan and Zamolodchikov, Alexander},
  year = {2022},
  journal = {Journal of High Energy Physics},
  volume = {2022},
  number = {8},
  pages = {57},
  doi = {10.1007/JHEP08(2022)057},
  eprint = {2203.11262},
    archivePrefix = {arXiv},
    primaryClass = {hep-th}
}

@article{IkhlefJacobsenSaleur2012,
  author = {Ikhlef, Yacine and Jacobsen, Jesper Lykke and Saleur, Hubert},
  title = {An integrable spin chain for the {$SL(2,\mathbb{R})/U(1)$} black hole sigma model},
  journal = {Physical Review Letters},
  volume = {108},
  number = {8},
  pages = {081601},
  year = {2012},
  doi = {10.1103/PhysRevLett.108.081601},
  eprint = {1109.1119},
  archivePrefix = {arXiv},
  primaryClass = {hep-th}
}

@article{FrahmSeel2014,
  author = {Frahm, Holger and Seel, Alexander},
  title = {The staggered six-vertex model: conformal invariance and corrections to scaling},
  journal = {Nuclear Physics B},
  volume = {879},
  pages = {382--406},
  year = {2014},
  doi = {10.1016/j.nuclphysb.2013.12.015},
  eprint = {1311.6911},
  archivePrefix = {arXiv},
  primaryClass = {cond-mat.stat-mech}
}

@article{BazhanovKotousovKovalLukyanov2021,
  author = {Bazhanov, Vladimir V. and Kotousov, Gleb A. and Koval, Sergii M. and Lukyanov, Sergei L.},
  title = {Scaling limit of the {$Z_2$} invariant inhomogeneous six-vertex model},
  journal = {Nuclear Physics B},
  volume = {965},
  pages = {115337},
  year = {2021},
  doi = {10.1016/j.nuclphysb.2021.115337},
  eprint = {2010.10613},
  archivePrefix = {arXiv},
  primaryClass = {math-ph}
}

@article{RobertsonJacobsenSaleur2021,
  author = {Robertson, Niall F. and Jacobsen, Jesper Lykke and Saleur, Hubert},
  title = {Lattice regularisation of a non-compact boundary conformal field theory},
  journal = {Journal of High Energy Physics},
  volume = {2021},
  number = {2},
  pages = {180},
  year = {2021},
  doi = {10.1007/JHEP02(2021)180},
  eprint = {2012.07757},
  archivePrefix = {arXiv},
  primaryClass = {hep-th}
}

@article{FrahmGehrmannKotousov2024,
  author = {Frahm, Holger and Gehrmann, Sascha and Kotousov, Gleb A.},
  title = {Scaling limit of the staggered six-vertex model with {$U_q(\mathfrak{sl}(2))$} invariant boundary conditions},
  journal = {SciPost Physics},
  volume = {16},
  number = {6},
  pages = {149},
  year = {2024},
  doi = {10.21468/SciPostPhys.16.6.149},
  eprint = {2312.11238},
  archivePrefix = {arXiv},
  primaryClass = {hep-th}
}

@article{ChepigaMila2017,
  title = {Excitation spectrum and density matrix renormalization group iterations},
  author = {Chepiga, Natalia and Mila, Fr{\'e}d{\'e}ric},
  journal = {Physical Review B},
  volume = {96},
  number = {5},
  pages = {054425},
  year = {2017},
  doi = {10.1103/PhysRevB.96.054425},
  eprint = {1705.05423},
  archivePrefix = {arXiv},
  primaryClass = {cond-mat.str-el}
}

@misc{supplement,
	note={{See the {Supplemental} {Material} at [URL will be inserted by the publisher] for finite-size correction analysis of projected amplitudes, {Hermitian} {Ising} and three-state {Potts} benchmark, {Yang--Lee} benchmark values and numerical details, and non-{Hermitian} five-state {Potts} Hamiltonian conventions and fit diagnostics, which includes Refs. [63-68].}}
}

@article{DateJimboMiwaOkado1987,
  author = {Date, Etsuro and Jimbo, Michio and Miwa, Tetsuji and Okado, Masato},
  title = {Automorphic properties of local height probabilities for integrable solid-on-solid models},
  journal = {Physical Review B},
  volume = {35},
  number = {4},
  pages = {2105--2107},
  year = {1987},
  doi = {10.1103/PhysRevB.35.2105}
}

@article{SaleurBauer1989,
  author = {Saleur, Hubert and Bauer, Michel},
  title = {On some relations between local height probabilities and conformal invariance},
  journal = {Nuclear Physics B},
  volume = {320},
  number = {3},
  pages = {591--624},
  year = {1989},
  doi = {10.1016/0550-3213(89)90014-X}
}

@article{Foda2018,
  author = {Foda, Omar},
  title = {Off-critical local height probabilities on a plane and critical partition functions on a cylinder},
  journal = {Nuclear Physics B},
  volume = {928},
  pages = {279--306},
  year = {2018},
  doi = {10.1016/j.nuclphysb.2018.01.011},
  eprint = {1711.03337},
  archivePrefix = {arXiv},
  primaryClass = {hep-th}
}

@article{CardyBulkRG2017,
  author = {Cardy, John},
  title = {Bulk Renormalization Group Flows and Boundary States in Conformal Field Theories},
  journal = {SciPost Physics},
  volume = {3},
  number = {2},
  pages = {011},
  year = {2017},
  doi = {10.21468/SciPostPhys.3.2.011},
  eprint = {1706.01568},
  archivePrefix = {arXiv},
  primaryClass = {hep-th}
}

@article{AresRajabpourViti2020,
  author = {Ares, Filiberto and Rajabpour, M. A. and Viti, Jacopo},
  title = {Scaling of the Formation Probabilities and Universal Boundary Entropies in the Quantum {XY} Spin Chain},
  journal = {Journal of Statistical Mechanics: Theory and Experiment},
  volume = {2020},
  number = {8},
  pages = {083111},
  year = {2020},
  doi = {10.1088/1742-5468/aba9d4},
  eprint = {2004.10606},
  archivePrefix = {arXiv},
  primaryClass = {cond-mat.stat-mech}
}

@article{LencsesVitiTakacs2019,
  author = {Lencs{\'e}s, M{\'a}t{\'e} and Viti, Jacopo and Tak{\'a}cs, G{\'a}bor},
  title = {Chiral entanglement in massive quantum field theories in 1+1 dimensions},
  journal = {Journal of High Energy Physics},
  volume = {2019},
  number = {1},
  pages = {177},
  year = {2019},
  doi = {10.1007/JHEP01(2019)177},
  eprint = {1811.06500},
  archivePrefix = {arXiv},
  primaryClass = {hep-th}
}

@article{Konechny2023,
  author = {Konechny, Anatoly},
  title = {{RG} boundaries and {Cardy's} variational ansatz for multiple perturbations},
  journal = {Journal of High Energy Physics},
  volume = {2023},
  number = {11},
  pages = {004},
  year = {2023},
  doi = {10.1007/JHEP11(2023)004},
  eprint = {2306.13719},
  archivePrefix = {arXiv},
  primaryClass = {hep-th}
}

@Article{10.21468/SciPostPhysCodeb.4,
	title={{The ITensor Software Library for Tensor Network Calculations}},
	author={Matthew Fishman and Steven R. White and E. Miles Stoudenmire},
	journal={SciPost Phys. Codebases},
	pages={4},
	year={2022},
	publisher={SciPost},
	doi={10.21468/SciPostPhysCodeb.4},
	url={https://scipost.org/10.21468/SciPostPhysCodeb.4},
}

@Article{10.21468/SciPostPhysCodeb.4-r0.3,
	title={{Codebase release 0.3 for ITensor}},
	author={Matthew Fishman and Steven R. White and E. Miles Stoudenmire},
	journal={SciPost Phys. Codebases},
	pages={4-r0.3},
	year={2022},
	publisher={SciPost},
	doi={10.21468/SciPostPhysCodeb.4-r0.3},
	url={https://scipost.org/10.21468/SciPostPhysCodeb.4-r0.3},
}

\begin{center}
{\large\bfseries END MATTER}
\end{center}

\setcounter{equation}{0}
\renewcommand{\theequation}{A\arabic{equation}}
\section*{Appendix~A: Continuum interpretation of projected coefficients}
We demonstrate how the lattice projected amplitude used in the main text is related to the universal BCFT data. Let
the conformal boundary state $a$ be expanded as
\begin{align}
  \lvert B_a\rangle
  =
  \sum_i B_a^{\,i}\lvert i\rangle\!\rangle ,
  \label{eq:app-boundary-state}
\end{align}
where $\lvert i\rangle\!\rangle$ denotes the Ishibashi state in bulk sector $i$~\cite{Ishibashi1989}.
Projecting the closed-channel cylinder onto sector $i$ gives
\begin{align}
  \mathcal{Z}_{a,i}^{\rm CFT}(\ell)
  &=
  \langle B_a\rvert
  e^{-\ell H_{\rm CFT}/2}
  \mathcal{P}_i
  e^{-\ell H_{\rm CFT}/2}
  \lvert B_a\rangle\notag\\
  &=
  e^{-\ell E_i^{\rm CFT}}
  G_{a,i} ,
  \label{eq:app-cft-projected-cylinder}
\end{align}
with
\begin{align}
  G_{a,i} = \left(B_a^{\,i}\right)^2 .
  \label{eq:app-G-B}
\end{align}
We note that by summing the projected amplitude $\mathcal{Z}_{a,i}^{\rm CFT}$, one recovers the usual cylinder partition function $\mathcal{Z}_{a}^{\rm CFT}=
\sum_i\mathcal{Z}_{a,i}^{\rm CFT}$.

To evaluate this projected partition function from a microscopic lattice model, we need a lattice
representative of the conformal boundary state $\left|B_a\right\rangle$. We take this representative to be 
an SRE
preparation $\left|\Phi\right\rangle$ whose long-distance limit is a regularized
conformal boundary state. 
This identification is part of the 
boundary-state framework: in
integrable RSOS/SOS models, off-critical local height probabilities and critical cylinder partition functions
with fixed boundary conditions were related in Refs.~\cite{DateJimboMiwaOkado1987,SaleurBauer1989} and
reviewed in Ref.~\cite{Foda2018}, while in continuum CFT Cardy's variational ansatz represents ground states
of relevant bulk perturbations by smeared conformal boundary states~\cite{CardyBulkRG2017}.

More explicitly, we use the Calabrese--Cardy regularized-boundary-state ansatz
~\cite{CalabreseCardy2006,CardyBulkRG2017},
\begin{align}
  \left|\Phi\right\rangle
  \propto
  e^{-\tau_\Phi H_{\rm CFT}}\left|B_a\right\rangle ,
  \label{eq:app-CC-ansatz}
\end{align}
where the nonuniversal smearing time $\tau_\Phi$ appears as a length scale. In Hermitian systems, this picture successfully describes the behavior of quantum quenches ~\cite{CalabreseCardy2006} and underlies overlap, formation-probability, and
partition-function-ratio extractions of boundary entropies and Cardy-state data
~\cite{Stephan2010,TangQMC2017,AresRajabpourViti2020}. Numerical tests and refinements of this ansatz include Refs.~\cite{LencsesVitiTakacs2019,Konechny2023}.

In non-Hermitian systems, the same principle should apply, but only if one 
implements
the left-right pairing underlying the infrared CFT.
In nonunitary BCFT, the bra boundary state is the linear dual defined by the CFT bilinear, or BPZ,
pairing~\cite{BelavinPolyakovZamolodchikov1984}, rather than the Hermitian conjugate of the ket boundary state.
Equivalently, recent pseudo-Hermitian/nonunitary BCFT constructions use dual Ishibashi states instead of
complex-conjugate coefficients~\cite{FukusumiKawamoto2026}.
In our formalism, we specify a left-right eigenstate pairing using symmetry of the non-Hermitian Hamiltonian. For each pairing protocol considered here, this symmetry provides a natural map from right to left eigenvectors. Once the pairing is specified, the dual preparation $\langle\widetilde{\Phi}|$ is determined by the right preparation $|\Phi\rangle$, as implemented in Eqs.~\eqref{eq:eta-dual-main} and \eqref{eq:complex-dual-main}.

With this pairing structure implemented, the lattice projected amplitude is the microscopic counterpart of the
projected closed-channel cylinder. Assume that the 
preparation 
of the right state $\left|\Phi\right\rangle$
and its normalized left dual $\langle\widetilde\Phi|$ realize
regularized boundary states,
\begin{align}
  \left|\Phi\right\rangle
  &\approx
  \mathcal N_{\Phi,R}(L)e^{-\tau_\Phi H_{\rm CFT}}\left|B_a\right\rangle_R,
  \notag\\
  \langle\widetilde\Phi|
  &\approx
  \mathcal N_{\Phi,L}(L)
  {}_L\!\langle\widetilde B_a|e^{-\widetilde\tau_\Phi H_{\rm CFT}} .
  \label{eq:app-dual-CC-ansatz}
\end{align}
Together with the biorthogonal projector
\begin{align}
    \frac{\left|\Psi_{i,R}\right\rangle\left\langle\Psi_{i,L}\right|}
  {\left\langle\Psi_{i,L}|\Psi_{i,R}\right\rangle}
  \approx \mathcal P_i ,
  \label{eq:app-P}
\end{align}
this gives
\begin{align}
  Z_{\Phi,i}(L)
  &\approx
  \mathcal N_{\Phi,L}(L)\mathcal N_{\Phi,R}(L)
  e^{-(\tau_\Phi +\widetilde\tau_\Phi )E_i^{\rm CFT}(L)}
  G_{a,i}.
\end{align}
Since $E_i^{\rm CFT}(L)=O(L^{-1})$, the smearing contributes only finite-size corrections to $-\log
Z_{\Phi,i}$. The remaining normalization factor is fixed by $\langle\widetilde{\Phi}|\Phi\rangle=1$ and therefore has the expression of a cylinder partition function $\mathcal{Z}_a^{\rm CFT}(\tau_\Phi+\widetilde\tau_\Phi)$. For the elementary Cardy boundary states considered here, the
leading open-channel multiplicity is unity; hence the normalization contributes only a nonuniversal extensive
term. We therefore obtain
\begin{equation}
  -\log Z_{\Phi,i}(L)
  \approx
  \alpha_\Phi L-\log G_{a(\Phi),i}+O(L^{-1}),
\end{equation}
which is the scaling form used in the main text. Further subleading finite-size corrections come from the
contributions of boundary and bulk irrelevant perturbations in the approximation in
Eqs.~\eqref{eq:app-dual-CC-ansatz} and \eqref{eq:app-P}, and are analyzed in the Supplemental Material~\cite{supplement}.

\section{Appendix~B: Analytic continuation of conformal loop model for the five-state Potts BCFT data}
\setcounter{equation}{0}
\renewcommand{\theequation}{B\arabic{equation}}
\label{app:potts-2BTL}

We summarize the BCFT data of the non-Hermitian five-state Potts model. The starting point is the
2BTL
partition function of the $Q$-state Potts
model~\cite{jacobsenConformalBoundaryLoop2008,dubailConformalTwoboundaryLoop2009}, whose analytic continuation
was used in Refs.~\cite{TangBoundary2025,VanderLinden2026} to describe open-chain spectra of the
non-Hermitian five-state Potts chain.  We use the same analytically continued boundary partition functions, but
read off the closed-channel coefficients, since the periodic projected overlap extracts
\begin{align}
    G_{k,\alpha}
  =
  \langle B_k|\alpha\rangle
  \langle \alpha|B_k\rangle .
\end{align}

The Potts partition function in the 2BTL formulation can be written as
\begin{align}
  Z_{\rm Potts}(q)
  =
  Z_{\rm loop}(q)
  +
  \left(Q_{12}-l_1l_2\right) Z_{l_1l_2}(q).
  \label{eq:app-2BTL-ZPotts}
\end{align}
The second term corrects the weight of noncontractible clusters touching the two boundaries.  For equal
$k$-mixed boundary conditions, in which both boundaries allow the same set of $k$ Potts colors, one has
\[
  Q_1=Q_2=Q_{12}=k,
  \qquad
  l_1=l_2=\frac{k}{\sqrt Q},
\]
and hence
\[
  Q_{12}-l_1l_2
  =
  k-\frac{k^2}{Q}.
\]

We use the Coulomb-gas parametrization
\begin{align}
  \sqrt Q = 2\cos\gamma,
  \qquad
  g_{\rm cg}=1-\frac{\gamma}{\pi}
  \label{eq:app-2BTL-cg}
\end{align}
A blob boundary condition allowing $Q_B=k$ Potts colors is parametrized by $r=r_k$, defined through
\begin{align}
  \frac{k}{\sqrt Q}
  =
  \frac{\sin[(r_k+1)\gamma]}{\sin(r_k\gamma)} .
  \label{eq:app-2BTL-r}
\end{align}

The closed-channel identity contribution comes from the physical loop term $Z_{\rm loop}$, with
$l=\sqrt Q$, or equivalently $\chi=\gamma$.  The $p=0$ sector gives
\begin{align}
  G_k^{\rm 2BTL}
  =
  G_{k,\mathbbm{1}}
  =
  g_k^2
  =
  (2g_{\rm cg})^{-1/2}
  \frac{\sin\gamma}{\sin^2(r_k\gamma)}
  \frac{
    \sin^2(r_k\gamma/g_{\rm cg})
  }{
    \sin(\gamma/g_{\rm cg})
  } .
  \label{eq:app-2BTL-master}
\end{align}
The correction term $Z_{l_1l_2}$ does not contribute to this identity projection, since it extracts the
$l_1l_2l^0$ coefficient and therefore shifts the closed-channel momentum sector away from the vacuum.

The same boundary-state expression gives the leading spin-sector projection.  The $Z_{l_1l_2}$ projection
sets $l=2\cos\chi=0$, i.e., $\chi=\pi/2$.  
In the first closed-channel family,
\begin{align}
  h_\alpha
  =
  \frac{1}{4g_{\rm cg}}
  \left[
    \left(p+\frac{\chi}{\pi}\right)^2
    -
    \left(\frac{\gamma}{\pi}\right)^2
  \right],
  \qquad p\in\mathbb Z ,
\end{align}
the two momenta $p=0,-1$ give the analytically continued Potts spin field,
\begin{align}
  h_\sigma
  =
  \frac{1}{4g_{\rm cg}}
  \left[
    \frac14
    -
    \left(\frac{\gamma}{\pi}\right)^2
  \right].
  \label{eq:app-2BTL-hsigma}
\end{align}
Taking the $l_1l_2$ coefficient of the corresponding closed-channel amplitude gives
\begin{align}
  A_k^{(\sigma)}
  =
  2(2g_{\rm cg})^{-1/2}
  \frac{
    \sin^2\!\left(
      \frac{r_k\gamma}{2g_{\rm cg}}
    \right)
  }{
    \cos\!\left(
      \frac{\gamma}{2g_{\rm cg}}
    \right)
  },
  \label{eq:app-2BTL-Ak-sigma}
\end{align}
where the factor of two comes from the two momenta $p=0,-1$. Therefore, the projection onto the full spin
multiplet is
\begin{align}
  \left.Z_{\rm Potts}^{(k,k)}\right|_{\sigma}
  =
  G_{k,\sigma}
  \frac{\tilde q^{-c/12}}{P(\tilde q^2)}
  \tilde q^{2h_\sigma},
  \quad
  G_{k,\sigma}
  =
  \left(k-\frac{k^2}{Q}\right)A_k^{(\sigma)} .
  \label{eq:app-2BTL-Ck-sigma}
\end{align}
Thus, the universal excited-to-ground boundary coefficient ratio is
\begin{align}
  R_{k,\sigma}
  =
  \frac{G_{k,\sigma}}{G_k^{\rm 2BTL}} .
  \label{eq:app-2BTL-Rk-sigma}
\end{align}
For $Q=5$, the spin sector is the four-dimensional $S_5$ vector multiplet, so
$R_{k,\sigma}$ should be compared with the numerical projection summed over the four nearly degenerate
first excited states~\footnote{One can also project the degenerate subspace to states with certain $S_5$ charges. 
While this charge-resolved projection is not needed for our purpose, 
it may be useful in more general applications.}.  For $k=5$, corresponding to the $S_5$-symmetric free boundary condition,
$G_{5,\sigma}=0$.

For the non-Hermitian five-state Potts fixed point, the analytic continuation beyond $Q=4$ makes $\gamma$
imaginary.  
We choose the branch
\begin{align}
  \gamma=-i\ell,
  \quad
  \ell=\log\varphi,
  \quad
  \varphi=\frac{1+\sqrt5}{2},
  \quad
  Q=5 .
  \label{eq:app-Q5-gamma}
\end{align}
The opposite sign of $\gamma$ gives the complex-conjugate fixed points.

For $Q\leq 4$, Eq.~\eqref{eq:app-2BTL-r} admits positive real solutions
\begin{align}
  r_k&= \frac{1}{\gamma} \tan^{-1}\left(\frac{\sin2\gamma}{k-1-\cos 2\gamma}\right)
\end{align}
that correspond to the cabling constructions of the boundary condition when $k\leq Q\in \mathbb{Z}$. Therefore, for $Q>4$, we simply adopt these solutions. Overall, for $k=1,\dots,5$,
\begin{align}
  r_1&=-2+\frac{i\pi}{\ell},\quad
  r_2=-1+\frac{i\pi}{2\ell},\notag\\
  r_3&=1+\frac{i\pi}{2\ell},\quad
  r_4=2,\quad
  r_5=1 .
  \label{eq:app-Q5-r-values}
\end{align}
Substitution of these values into Eqs.~\eqref{eq:app-2BTL-master} and
\eqref{eq:app-2BTL-Rk-sigma} gives the boundary coefficients and the spin-sector
excited-to-ground ratios used for comparison with the projected overlaps. We note that the choice of $r_1$ is different from that in Ref.~\cite{TangBoundary2025}, where they take the real solution $r=-2$ and encounter additional characters in the open-string partition functions that are absent in the open chain spectrum. One can show that with $r_1$ provided in Eq.~\eqref{eq:app-Q5-r-values}, those characters disappear and the low-lying excited state coincides with the 2BTL partition functions.

\clearpage

\appendix

\begin{widetext}

\begin{center}
\textbf{\large Supplemental Material for ``Extracting Boundary Conformal Data from \\ Periodic Non-Hermitian Critical Chains''}
\end{center}

\setcounter{subsection}{0}
\setcounter{equation}{0}
\setcounter{figure}{0}
\renewcommand{\theequation}{S\arabic{equation}}
\renewcommand{\thefigure}{S\arabic{figure}}
\renewcommand{\thetable}{S\arabic{table}}
\setcounter{table}{0}

\let\addcontentsline\origaddcontentsline
\tableofcontents

\vspace{2\baselineskip} 

\section{A.~Finite-size corrections to the projected partition function}
\setcounter{equation}{0}
\renewcommand{\theequation}{A.\arabic{equation}}

In this section, we summarize the expected finite-size correction structure of the projected amplitudes. The purpose is not to determine all model-dependent amplitudes, but to
organize the powers that can appear in the extrapolation. We first discuss the general origin of
the corrections and then state the fitting conventions used for the Yang--Lee and five-state Potts analyses in
this work.

\subsection{I.~Leading projected-cylinder scaling}

The continuum interpretation in Appendix~A identifies the lattice projected amplitude with a regularized
closed-channel boundary cylinder. For an
SRE microscopic preparation and its compatible normalized left
dual, one may write schematically
\begin{align}
  \left|\Phi\right\rangle
  &\approx
  \mathcal N_{\Phi,R}(L)\,
  e^{-\tau_\Phi  H_{\rm CFT}}
  \left|B_a\right\rangle_R,
  \notag\\
  \langle\widetilde\Phi|
  &\approx
  \mathcal N_{\Phi,L}(L)\,
  {}_L\!\langle\widetilde B_a|
  e^{-\widetilde\tau_\Phi  H_{\rm CFT}} .
  \label{eq:fss-regularized-boundary}
\end{align}
Here $\tau_\Phi ,\widetilde\tau_\Phi $ are nonuniversal microscopic extrapolation lengths, while $\mathcal
N_{\Phi,L}\mathcal N_{\Phi,R}$ is fixed by the paired normalization $\langle\widetilde\Phi |\Phi \rangle=1$.
Projecting onto the periodic sector $i$ then gives
\begin{align}
  Z_{\Phi,i}(L)
  &\approx
  \mathcal N_{\Phi,L}(L)\mathcal N_{\Phi,R}(L)\,
  e^{-(\tau_\Phi +\widetilde\tau_\Phi )E_i^{\rm CFT}(L)}
  G_{a,i}.
  \label{eq:fss-leading-Z}
\end{align}
The same-boundary annulus entering the normalization has a leading open-channel multiplicity equal to one for
the elementary Cardy sectors considered in the main text. Consequently the normalization factor contributes
only a nonuniversal extensive term to $-\log Z_{\Phi,i}$. Since $E_i^{\rm CFT}(L)=O(L^{-1})$, the smearing
factor in Eq.~\eqref{eq:fss-leading-Z} produces the leading universal finite-size correction of order $1/L$.
Thus
\begin{equation}
  -\log Z_{\Phi,i}(L)
  =
  \alpha_\Phi  L
  -
  \log G_{a(\Phi),i}
  +
  \frac{\beta_\Phi}{L}+\cdots.
  \label{eq:fss-leading-logZ}
\end{equation}
The ratio
\begin{align}
  R_{\Phi,i}(L)=Z_{\Phi,i}(L)/Z_{\Phi,\iota_g}(L)
\end{align}
cancels the extensive term but generally retains $1/L$ corrections, because the smearing contribution depends
on the CFT energy of the projected sector. We take Eq.~\eqref{eq:fss-leading-logZ} as the general form and
discuss the possible finite-size corrections in addition to this always-present leading part. In the following, we will write $a=a(\Phi)$ for simplicity.

\subsection{II.~Boundary perturbations}

The microscopic preparation state need not be exactly the smeared Cardy state in
Eq.~\eqref{eq:fss-regularized-boundary}. Its infrared expansion may include irrelevant boundary perturbations
at the two regularized boundaries,
\begin{align}
  \left|\Phi\right\rangle
  \approx
  \mathcal N_{\Phi,R}(L)
  e^{-\tau_\Phi  H_{\rm CFT}}
  \left[
    1+\sum_{\mu}g_{\mu}^{R}L^{1-h_\mu}\mathcal O_{\mu}^{R}
    +\cdots
  \right]\left|B_a\right\rangle_R ,
  \label{eq:fss-boundary-perturbation}
\end{align}
and analogously for the normalized left dual. Here $h_\mu>1$ is the scaling dimension of a boundary irrelevant
operator $\mathcal{O}^{R/L}_\mu$ allowed by the microscopic symmetries and by the corresponding boundary
condition. To first order, such a boundary perturbation contributes
\begin{equation}
  \delta_{\rm bdy}
  \bigl[-\log Z_{\Phi,i}(L)\bigr]
  =
  \sum_{\mu} c_{\mu,\Phi,i}^{\rm bdy}\,L^{1-h_\mu}
  +\cdots .
  \label{eq:fss-boundary-power}
\end{equation}
Repeated insertions and the logarithm of the amplitude generate sums of such powers. 
In particular, the energy-momentum tensor has $h=2$ and gives a $1/L$~\cite{FSCYifan} term that enters the third term in the general expression~\eqref{eq:fss-leading-logZ}.
Its square and higher boundary descendants can generate
integer corrections such as $1/L^2$ and $1/L^3$ as the leading finite-size corrections. If a different
boundary irrelevant operator with dimension $1<h_b<2$ is the leading allowed correction, the first
boundary-induced power is instead $L^{1-h_b}$.

\subsection{III.~Bulk irrelevant perturbations and eigenstate corrections}

There are also corrections from the fact that the microscopic periodic-chain Hamiltonian is not exactly the
CFT Hamiltonian. Near criticality one may write
\begin{equation}
  H_{\rm lat}
  =
  H_{\rm CFT}
  +
  \sum_{\nu} g_{\nu}
  \int_0^L dx\,\Phi_\nu(x)
  +\cdots ,
  \qquad
  x_\nu>2 ,
  \label{eq:fss-bulk-irrelevant}
\end{equation}
where $x_\nu$ is the bulk scaling dimension of an irrelevant field, and we do not explicitly include the
nonuniversal energy density term, as it only shifts the spectrum constantly. The integrated perturbation has
matrix elements scaling as $L^{1-x_\nu}$. It therefore shifts finite-size energies as
\begin{equation}
  \delta E_i(L)\sim L^{1-x_\nu}.
  \label{eq:fss-energy-correction}
\end{equation}
When inserted into the smearing factor of Eq.~\eqref{eq:fss-leading-Z}, this produces a correction to $-\log
Z_{\Phi,i}$ of order $L^{1-x_\nu}$.

Bulk perturbations also modify the finite-size eigenstates appearing in the projected overlap. First-order
perturbation theory gives schematically
\begin{align}
  \left|\Psi_{i,R}^{\rm lat}\right\rangle
  =
  \left|i\right\rangle_R
  +
  \sum_{j\ne i,\nu}
  \frac{
    g_\nu\,
    {}_L\!\left\langle j\right|
    \int_0^L dx\,\Phi_\nu(x)
    \left|i\right\rangle_R
  }{
    E_i^{\rm CFT}(L)-E_j^{\rm CFT}(L)
  }
  \left|j\right\rangle_R
  +\cdots .
  \label{eq:fss-eigenstate-perturbation}
\end{align}
Since $E_i^{\rm CFT}-E_j^{\rm CFT}=O(L^{-1})$, this correction scales as
\begin{equation}
  \delta\left|\Psi_i\right\rangle
  \sim
  L^{2-x_\nu}.
  \label{eq:fss-eigenstate-power}
\end{equation}
Thus a bulk irrelevant field of dimension $x_\nu$ can enter the projected overlap through two different
powers:
\begin{align}
  L^{2-x_\nu}
  \quad
  \text{from eigenstate mixing,}
  \qquad
  L^{1-x_\nu}
  \quad
  \text{from the energy shift in the smearing factor.}
\end{align}
For the common scalar irrelevant perturbation $T\bar T$ with $x_\nu=4$, these powers are $1/L^2$ and $1/L^3$,
respectively.

\subsection{IV.~General effective expansion}

Combining the regularized-boundary smearing, boundary irrelevant perturbations, bulk eigenstate corrections,
and bulk energy corrections
leads to the following effective expansion of the projected amplitude,
\begin{align}
  -\log Z_{\Phi,i}(L)
  &=
  \alpha_\Phi  L
  -
  \log G_{a,i}
  +
  \frac{A_{\Phi,i}}{L}
  \notag\\
  &\quad
  +
  \sum_{\mu} B_{\mu,\Phi,i} L^{1-h_\mu}
  +
  \sum_{\nu} C_{\nu,i} L^{2-x_\nu}
  +
  \sum_{\nu} D_{\nu,i} L^{1-x_\nu}
  +\cdots .
  \label{eq:fss-general-expansion}
\end{align}
The displayed $1/L$ term includes the universal smearing correction and any boundary stress-tensor
contribution. The remaining sums should be understood as the leading allowed corrections determined by the
boundary and bulk operator contents of the microscopic realization. Products of corrections and the expansion
of the logarithm generate additional powers obtained by adding the exponents appearing in
Eq.~\eqref{eq:fss-general-expansion}.

For sector ratios, the nonuniversal extensive term cancels:
\begin{align}
  R_{\Phi,i}(L)
  =
  \frac{G_{a,i}}{G_{a,\iota_g}}
  +
  \frac{A_{\Phi,i}}{L}
  +
  \sum_{\mu} B_{\mu,\Phi,i} L^{1-h_\mu}
  +
  \sum_{\nu} C_{\nu,i} L^{2-x_\nu}
  +\cdots,
  \label{eq:fss-ratio-general}
\end{align}
where the leading corrections come directly from the combination of terms in Eq.~\eqref{eq:fss-general-expansion} for $Z_{\Phi,i}$ and $Z_{\Phi,\iota_g}$, and we only keep the leading term from the bulk perturbation.

In practice, a finite-size data set cannot determine all powers independently. We therefore use the
field-theory power counting to choose a minimal stable fit form and treat the remaining correction terms
through fit-window variations.

\subsection{V.~Fit choices used in the main analysis}

For the Yang--Lee projected amplitudes, we use the following two-correction fit:
\begin{equation}
  -\log Z_{\Phi,\phi}(L)
  =
  \alpha_\Phi  L
  -
  \log G_a^{\rm num}
  +
  \frac{A_\Phi}{L}
  +
  \frac{B_\Phi}{L^2},
  \label{eq:fss-YL-log-fit}
\end{equation}
and
\begin{equation}
  R_{\Phi,\mathbbm{1}}(L)
  \equiv\frac{Z_{\Phi,\mathbbm{1}}(L)}{Z_{\Phi,\phi}(L)}
 = R_{a,\mathbbm{1}}{(\infty)}
  +
  \frac{A^r_\Phi}{L}
  +
  \frac{B^r_\Phi}{L^2}.
  \label{eq:fss-YL-ratio-fit}
\end{equation}
We note that, as we only discuss the boundary condition corresponding to the identity operator in this model,
the irrelevant boundary operators can only be those in the identity family.
Bulk irrelevant perturbations of the Yang--Lee lattice realization were analyzed in
Ref.~\cite{xu2022}; the least irrelevant scalar perturbation is
identified as $T\bar T$.
As discussed above, such a bulk perturbation contributes to the projected overlap 
through finite-size eigenstate mixing at order $1/L^2$.
Equation~\eqref{eq:fss-YL-ratio-fit} therefore captures the leading expected analytic corrections for the available
Yang--Lee data.

For the non-Hermitian five-state Potts calculation, the finite-size window is more limited and the extrapolation is
performed for complex quantities. Adding higher correction powers leads to unstable multi-parameter fits and
strong correlations between the real and imaginary parts of the intercept. We therefore use the minimal
complex extrapolation
\begin{equation}
  -\log Z_k(L)
  =
  \alpha_k L
  -
  \log G_k^{\rm num}
  +
  \frac{A_k}{L},
  \label{eq:fss-potts-log-fit}
\end{equation}
and display the finite-size trajectories rather than claiming a detailed correction expansion. For the
first-excited-subspace diagnosis, we use the normalization-independent ratio
\begin{equation}
  R_{k,{\rm ex}}(L)
  \equiv
  \frac{Z_{k,{\rm ex}}(L)}{Z_k(L)}
  =
  R_{k,{\rm ex}}{(\infty)}
  +
  \frac{A_{k,{\rm ex}}}{L}.
  \label{eq:fss-potts-ratio-fit}
\end{equation}
The robustness of the Potts result is judged from the stability of the ground-sector intercepts and from this
excited-sector ratio diagnostic.

\section{B.~Benchmark for Hermitian systems}
\setcounter{equation}{0}
\renewcommand{\theequation}{B.\arabic{equation}}
In this section, we benchmark the projected-amplitude method in Hermitian critical chains, where the ordinary
Hermitian inner product gives the appropriate left-right pairing, and the expected boundary coefficients are well established.  
We use the transverse-field Ising (TF-Ising) chain and the three-state Potts chain, with
the Hamiltonians
\begin{align}
    H_{\rm Ising}=-\sum_{j=1}^L(\sigma_j^z\sigma_{j+1}^z+\lambda \sigma^x_j)
\end{align}
and
\begin{align}
    H_{\text{3-state}}=-\sum_j\left(R_j^{\dagger} R_{j+1}+R_j R_{j+1}^{\dagger}+ \lambda M_j+\lambda M_j^{\dagger} \right),
\end{align}
where
\begin{align}
R=\left(\begin{array}{ccc}
e^{2 \pi i / 3} & 0 & 0 \\
0 & e^{4 \pi i / 3} & 0 \\
0 & 0 & 1
\end{array}\right)\ ,\quad
    M=\left(\begin{array}{ccc}
0 & 1 & 0 \\
0 & 0 & 1 \\
1 & 0 & 0
\end{array}\right) .
\end{align}
Both models are critical at $\lambda=1$ and are described by the $M(3,4)$ and $M(5,6)$ minimal models,
respectively.

The conformal boundary conditions in these models can be represented microscopically by restricting the allowed
local spin values in the preparation state, in the same spirit as the Potts preparations used in the main
text.  For the critical Ising model, the elementary conformal boundary conditions are fixed $(+)$, fixed $(-)$,
and free~\cite{CardyBCFT2006}.  Since the two fixed boundaries are related by the global $\mathbb{Z}_2$ symmetry
and have the same $g$-factor, we keep only one fixed representative.  For the three-state Potts model, the
conformal boundary conditions include fixed, mixed, free, and ``new'' boundaries~\cite{AffleckOshikawaSaleur1998}.
The fixed, mixed, and free boundaries have direct product-state representatives; the microscopic realization of
the new boundary is less obvious~\cite{chepigaCriticalPropertiesQuantum2022}, so it is not included in this
benchmark.

The preparation states are
\begin{align}
    \left|\Psi_k\right\rangle\equiv\bigotimes_{j=1}^{L}
\frac{1}{\sqrt{k}}
\sum_{n=0}^{k-1}|n\rangle_j ,
\label{eq:supp_prep_Her}
\end{align}
where $k=1,\ldots,Q$, and $Q$ denotes the number of local states.
For the Ising model, $Q=2$, so $k=1$ and $k=2$ correspond to the
fixed and free preparations, respectively. For the three-state Potts
model, $Q=3$, and the choices $k=1,2,3$ correspond to the fixed,
mixed, and free preparations. In the Hermitian case, the compatible
left preparation is simply the Hermitian conjugate of the right
preparation.
The exact BCFT values used for the residuals in Fig.~\ref{fig:Hermitian} are
\begin{align}
  g_{\uparrow}=\frac{1}{\sqrt{2}},
  \qquad
  g_f=1,
  \label{eq:supp-ising-g-values}
\end{align}
for the Ising chain, and
\begin{align}
  g_A=N,
  \qquad
  g_{AB}=N\lambda^2,
  \qquad
  g_f=\sqrt{3}\,N,
  \label{eq:supp-potts3-g-values}
\end{align}
for the three-state Potts chain, where
$\lambda^2=(\sqrt{5}+1)/2$ and
$N^4=(5-\sqrt{5})/30$~\cite{AffleckOshikawaSaleur1998}.

The resulting projected coefficients are shown in Fig.~\ref{fig:Hermitian}.  
In both models, the extracted
boundary $g$-factors converge to the BCFT values with errors at the level of the third significant digit,
providing a Hermitian check of the projected-partition-function method.
\begin{figure}[t]
  \centering
  \includegraphics[width=\columnwidth]{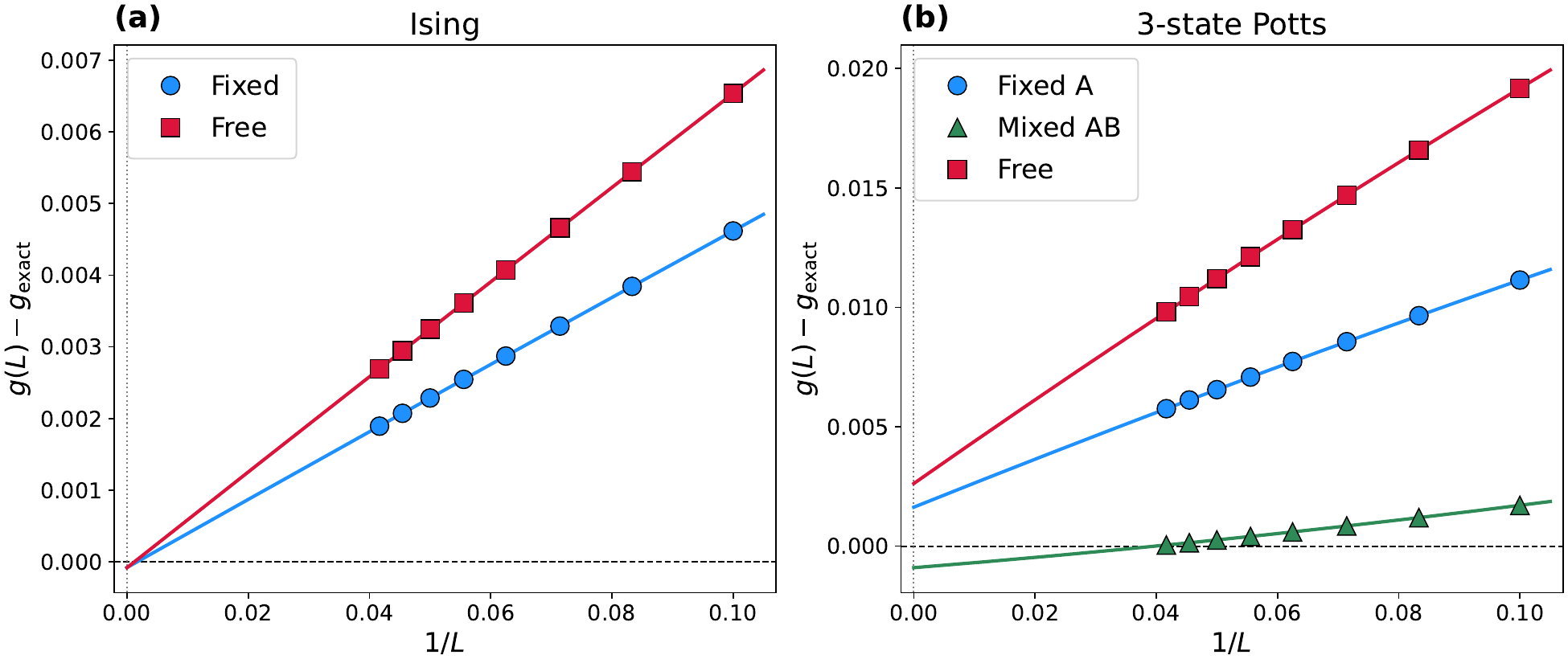}
  \caption{
  Finite-size convergence of extracted $g$-factors in the (a)~TF-Ising and (b)~three-state Potts models.  Residuals
  $g(L)-g_{\rm exact}$ are plotted for the preparations defined in Eq.~\eqref{eq:supp_prep_Her}.  Finite-size
  extrapolations use the $1/L+1/L^2$ ansatz.
  }
  \label{fig:Hermitian}
\end{figure}

\section{C.~Yang--Lee BCFT benchmarks and numerical details}
\setcounter{equation}{0}
\renewcommand{\theequation}{C.\arabic{equation}}
\subsection{I.~BCFT benchmark values}
\label{sec:supp-YL-BCFT}
We summarize the Yang--Lee BCFT data
for comparison with the results in the main text.
The Yang--Lee CFT is the
nonunitary minimal model $M(2,5)$, with central charge $c=-22/5$ and two primaries $\mathbbm{1}$ and $\phi$,
whose conformal weights are
\begin{align}
  h_{\mathbbm{1}}=0,
  \qquad
  h_{\phi}=-\frac15 .
\end{align}
The physical finite-size ground state belongs to the sector
\begin{align}
  o=\phi ,
\end{align}
because $\phi$ has the lowest conformal weight. In the basis $(\mathbbm{1},\phi)$, the modular $S$-matrix is
\begin{align}
  S
  =
  \frac{2}{\sqrt5}
  \begin{pmatrix}
    -\sin(2\pi/5) & \sin(\pi/5) \\
     \sin(\pi/5) & \sin(2\pi/5)
  \end{pmatrix}.
  \label{eq:app-YL-S-matrix}
\end{align}

For a diagonal RCFT, Cardy's construction of boundary states still applies~\cite{CardyBoundary1989}, and the
boundary coefficients are
\begin{align}
  B_a^{\,i}
  =
  \frac{S_{ai}}{\sqrt{S_{\mathbbm{1} i}}},
  \qquad
  G_{a,i}
  =
  \left(B_a^{\,i}\right)^2
  =
  \frac{S_{ai}^{\,2}}{S_{\mathbbm{1} i}} .
  \label{eq:app-Cardy-coefficients}
\end{align}
A key distinction from the unitary case is that the boundary coefficients $B_a^{\,i}$ are not necessarily real,
and the projected coefficients $G_{a,i}$ can be negative, which is observed in the main text. Overall, with
this convention, related universal data are
\begin{align}
  G_{\mathbbm{1},\phi}
  &=
  g_{\mathbbm{1}}^{\,2}
  =
  S_{\mathbbm{1}\phi}
  =
  \sqrt{\frac{5-\sqrt5}{10}},
  \qquad
  g_{\mathbbm{1}}
  =
  \left(\frac{5-\sqrt5}{10}\right)^{1/4}
  ,
  \notag\\
  G_{\mathbbm{1},\mathbbm{1}}
  &=
  S_{\mathbbm{1}\mathbbm{1}}
  =
  -\sqrt{\frac{5+\sqrt5}{10}}
  ,
  \label{eq:app-YL-minimal-g}
\end{align}
which are the values we compare with in the main text.

\subsection{II.~Critical points and details of numerical simulations}
\label{sec:supp-YL-numerics}

In this section, we provide details of the numerical simulation for the Yang--Lee BCFT in the main text.
Results presented for this model are obtained by exact diagonalization. The Yang--Lee{}
projected-partition-function data are not computed by retuning each finite system to a size-dependent
exceptional point. Instead, for each transverse-field coupling $h_x\equiv\lambda$, finite-size transition
points $h_c(L;\lambda)$ are evaluated when the ground state energy turns from real to complex and extrapolated
to a thermodynamic critical field $h_c^{(\infty)}(\lambda)$.

The extrapolation is performed by fitting the finite-size critical fields to the form
\begin{equation}
  h_c(L;\lambda)
  =
  h_c^{(\infty)}(\lambda)
  +A L^{-2.4}+B L^{-4.4}+C L^{-4.8}+\cdots.
  \label{eq:supp-YL-hc-fit}
\end{equation}
The powers in Eq.~\eqref{eq:supp-YL-hc-fit} are motivated by a perturbed-CFT argument. In fact, one may
consider the following effective Hamiltonian for the finite-size system
\begin{align}
  H_{\text{eff}} = H_{CFT} + \lambda \int \mathrm{d}x \, \phi(x) + \alpha \int \mathrm{d}x \, T\bar{T}(x),
\end{align}
where $H_{CFT}$ is the Yang--Lee CFT Hamiltonian, $\phi$ is the relevant perturbation with scaling dimension
$2h_\phi=-\frac25$, and $T\bar{T}$ is the leading irrelevant perturbation with scaling dimension $4$~\cite{xu2022}. For a
system with length $L$, the energy gap can be estimated by the scaling analysis as
\begin{align}
  \Delta E(L) = \frac{1}{L} \mathcal{F}(u, v),
\end{align}
where $u = \lambda L^{2-2h_\phi}$ and $v = \alpha L^{2-4}$ are scalar variables associated with the
perturbations. The critical point is determined by the condition $\Delta E(L) = 0$, so the critical condition
is $\mathcal{F}(u_c,v)=0$. Assuming that $u_c$ can be expanded as a power series of $v$, we have
\begin{align}
  u_c = u_0 + u_1 v + O(v^2).
\end{align}
Substituting the scaling variables into the critical condition, we have
\begin{align}
  \lambda_c L^{2.4} = u_0 + u_1 \alpha L^{-2} + O(L^{-4}).
  \label{eq:supp-YL-lambdac}
\end{align}
In the real lattice model, the microscopic field $h_z$ is related to the conformal perturbation $\lambda$ by
an analytical nonlinear map, so the critical field $h_c(L)$ can be expanded as a power series of $\lambda_c$
as
\begin{align}
  h_c(L) = h_c^{(\infty)} + a_1 \lambda_c + a_2 \lambda_c^2 + O(\lambda_c^3).
\end{align}
Substituting Eq.~\eqref{eq:supp-YL-lambdac} into the above expansion, we arrive at the fitting form in
Eq.~\eqref{eq:supp-YL-hc-fit}. The detailed fitting results are summarized in Table~\ref{tab:hc}.
\begin{table}[tb]
\centering
\caption{Finite-size Yang--Lee exceptional points $h_c(L;\lambda)$, obtained by exact diagonalization, and
the extrapolated critical field $h_c^{(\infty)}(\lambda)$ for $L \to \infty$. 
The finite-size entries are
truncated at the numerical precision of the bisection search used to locate the exceptional points.
The parentheses in the $L \to \infty$ column denote the uncertainty of the finite-size extrapolation.}
\label{tab:hc}
\resizebox{\textwidth}{!}{%
\begin{tabular}{ccccccccc}
\toprule
\multicolumn{1}{c}{$h_x=\lambda$} &
\multicolumn{7}{c}{$L$} &
\multicolumn{1}{c}{$L\to\infty$} \\
\cmidrule(lr){2-8}
& 10 & 12 & 14 & 16 & 18 & 20 & 22 & $h_c^{(\infty)}$ \\
\midrule
2.0  & 0.32903544977 & 0.32338105419 & 0.32009475154 & 0.31803991863 & 0.31668123066 & 0.31574231815 & 0.31506990475 & $0.3124029(3)$ \\
3.0  & 0.89965188944 & 0.89402692330 & 0.89081429898 & 0.88882923005 & 0.88752762734 & 0.88663365086 & 0.88599635383 & $0.8834983(1)$ \\
4.0  & 1.57505330308 & 1.56932743241 & 1.56607420069 & 1.56407088458 & 1.56276039474 & 1.56186182547 & 1.56122205060 & $1.55872195(5)$ \\
5.0  & 2.30917588986 & 2.30330512436 & 2.29997768138 & 2.29793188259 & 2.29659503533 & 2.29567909175 & 2.29499213406 & $2.29238(5)$ \\
\bottomrule
\end{tabular}%
}
\end{table}

The projected partition function $Z_{\Phi,i}^{(\eta)}(L)$ is evaluated with the three preparation states
defined in the main text, with the same system-size window
as the critical field extrapolation ($L=10\sim22$).
In the main text, we only present the results for $\lambda=4.0$, as it has the highest accuracy in the critical
field extrapolation. 
The results for $\lambda=2.0$ and $\lambda=3.0$ are summarized in Table~\ref{tab:Z}. 
We find that from $\lambda=5.0$ the critical field extrapolation becomes less reliable and the results become
noisy, so we do not present the results for $\lambda\geq 5.0$ in this work.

\begin{table}[tb]
\centering
\caption{Yang--Lee projected-partition-function benchmarks for the two additional microscopic critical realizations.
The fitting form, fitting window, and uncertainty convention are the same as in the main text:
$L=10,12,\ldots,22$ is used for the central $1/L+1/L^2$ extrapolation, and the parentheses denote the
combined finite-size-window and $h_c^{(\infty)}$ uncertainty.}
\label{tab:Z}
\begin{tabular}{ccccc}
\toprule
prep. &
$\lambda$ &
$g_{\mathbbm{1}}^{\rm num}$ &
$R(\infty)$ &
$\epsilon_\eta/\epsilon_0$ \\
\midrule
$X$ &
2 &
$0.7245(11)$ &
$-1.6293(22)$ &
$10^{-12}/-$ \\
$Z$ &
2 &
-- &
$-1.621(6)$ &
$10^{-12}/10^{-1}$ \\
$\mathrm{Rand}$ &
2 &
-- &
$-1.640(11)$ &
$10^{-12}/10^{-1}$ \\
\midrule
$X$ &
3 &
$0.72532(20)$ &
$-1.6241(10)$ &
$10^{-10}/-$ \\
$Z$ &
3 &
-- &
$-1.61783(92)$ &
$10^{-10}/10^{-1}$ \\
$\mathrm{Rand}$ &
3 &
-- &
$-1.6246(34)$ &
$10^{-10}/10^{-1}$ \\
\midrule
BCFT &
-- &
$0.725073$ &
$-1.618034$ &
-- \\
\bottomrule
\end{tabular}
\end{table}

\section{D.~Numerical details of the non-Hermitian five-state Potts model}
\setcounter{equation}{0}
\renewcommand{\theequation}{D.\arabic{equation}}

In this section, we provide the microscopic convention and finite-size fitting details for the non-Hermitian
five-state Potts calculation used in the main text.  The local Hilbert space is spanned by
$\{\left|n\right\rangle\}_{n=0}^{4}$.  We use the phase and shift operators
\begin{align}
  \sigma\left|n\right\rangle
  =
  \omega^n \left|n\right\rangle,
  \qquad
  \tau\left|n\right\rangle
  =
  \left|n+1\ {\rm mod}\ 5\right\rangle,
  \qquad
  \omega=e^{2\pi i/5},
\end{align}
following the convention of Refs.~\cite{TangBulk2024,VanderLinden2026}.  The periodic-chain Hamiltonian is
\begin{align}
  H_{\rm Potts}
  =
  H_0+\lambda H_1 ,
  \label{eq:supp-potts-H}
\end{align}
where
\begin{align}
  H_0
  =
  -
  \sum_{j=1}^{L}
  \sum_{m=1}^{4}
  \left[
    \left(\sigma_j^\dagger \sigma_{j+1}\right)^m
    +
    \tau_j^m
  \right],
  \label{eq:supp-potts-H0}
\end{align}
and
\begin{align}
  H_1
  =
  \sum_{j=1}^{L}
  \sum_{m,n=1}^{4}
  \Bigl[
    \left(\tau_j^m+\tau_{j+1}^m\right)
    \left(\sigma_j^\dagger \sigma_{j+1}\right)^n
    +
    \left(\sigma_j^\dagger \sigma_{j+1}\right)^m
    \left(\tau_j^n+\tau_{j+1}^n\right)
  \Bigr].
  \label{eq:supp-potts-H1}
\end{align}
The boundary condition is periodic, so $j+1$ is interpreted modulo $L$.  The calculations use $J=h=1$ and the complex self-dual fixed-point coupling
$\lambda=0.0788+0.0603i$, consistent with the value quoted in the main text.  The Hamiltonian is complex
symmetric in this basis.  Therefore the compatible left-right pairing used in the projected amplitude is the prescription
$\left\langle \Psi_L\right|=\left\langle \mathcal K\Psi_R\right|$ and
$\langle \widetilde\Phi_k|=\left\langle\Phi_k\right|$ for the real preparations
of Eq.~\eqref{eq:potts-preparation-family}.

To reach the system sizes used below, we compute the right ground-state MPSs using a two-site finite-system density-matrix renormalization-group (DMRG) algorithm. At each local update, the non-Hermitian effective eigenvalue problem is solved with an Arnoldi eigensolver, in place of the Lanczos procedure commonly used for Hermitian Hamiltonians. Because the usual variational monotonicity of Hermitian DMRG is absent in the non-Hermitian setting, convergence is not guaranteed a priori. We therefore assess convergence directly by increasing the MPS bond dimension and confirming that the targeted eigenvalue and the projected quantities entering our analysis change only negligibly. All DMRG calculations were performed using the \textsc{ITensor} library~\cite{10.21468/SciPostPhysCodeb.4,*10.21468/SciPostPhysCodeb.4-r0.3}.

The right ground-state MPS data used in the extrapolation are obtained for
\begin{align}
  L=8,16,18,20,22,24 .
\end{align}
For the finite-size extrapolation, we compare two simple ansatzes, one with a single $1/L$ correction and the other
with an additional $1/L^2$ correction.  The former is used for the central values in the main text, while the
latter is used as a stability diagnostic.
For a preparation $\Phi_k$ with $k=1,\ldots,5$, the finite-size projected amplitude is evaluated as
\begin{align}
  Z_k(L)
  =
  \frac{
  \langle \Phi_k|\Psi_R\rangle\langle \Psi_L|\Phi_k\rangle
  }{
  \langle \Psi_L|\Psi_R\rangle
  },
  \label{eq:supp-potts-Zraw}
\end{align}
where $\Phi_k$ is normalized as in Eq.~\eqref{eq:potts-preparation-family}.  Equivalently, in the raw
unnormalized product-state implementation, one divides by the factor $k^L$.  We then fit
\begin{align}
  -\log Z_k(L)
  =
  \alpha_k L-\log G_k^{\rm num}+\frac{A_k}{L},
  \qquad
  g_k^{\rm num}=\sqrt{G_k^{\rm num}}.
  \label{eq:supp-potts-fit}
\end{align}
The uncertainty quoted in Table~\ref{tab:potts-G-family} is obtained by
propagating the residual covariance of the unweighted complex least-squares fit in
Eq.~\eqref{eq:supp-potts-fit} to $g_k=\sqrt{G_k}$.  This procedure gives the values shown in the main
text.  Figure~\ref{fig:supp-potts-ground-g-fit} shows the corresponding finite-size trajectories of
$g_k=\sqrt{G_k}$ and compares the central $1/L$ fits with the $1/L+1/L^2$ fit ansatz.  In the ground sector
the $1/L$ extrapolation gives the more stable overall benchmark: in particular,
$G_{\rm free}/G_{\rm fixed}=5.00109+0.00016i$, in excellent agreement with the Kramers--Wannier prediction
$5$.  The additional $1/L^2$ parameter reduces the residual but is poorly
constrained over the available size window and produces appreciable
shifts of the extrapolated intercepts. We therefore use the minimal
$1/L$ fit for the central values and treat the two-correction fit as a
stability diagnostic.

\begin{figure}[tb]
  \centering
  \includegraphics[width=0.92\textwidth]{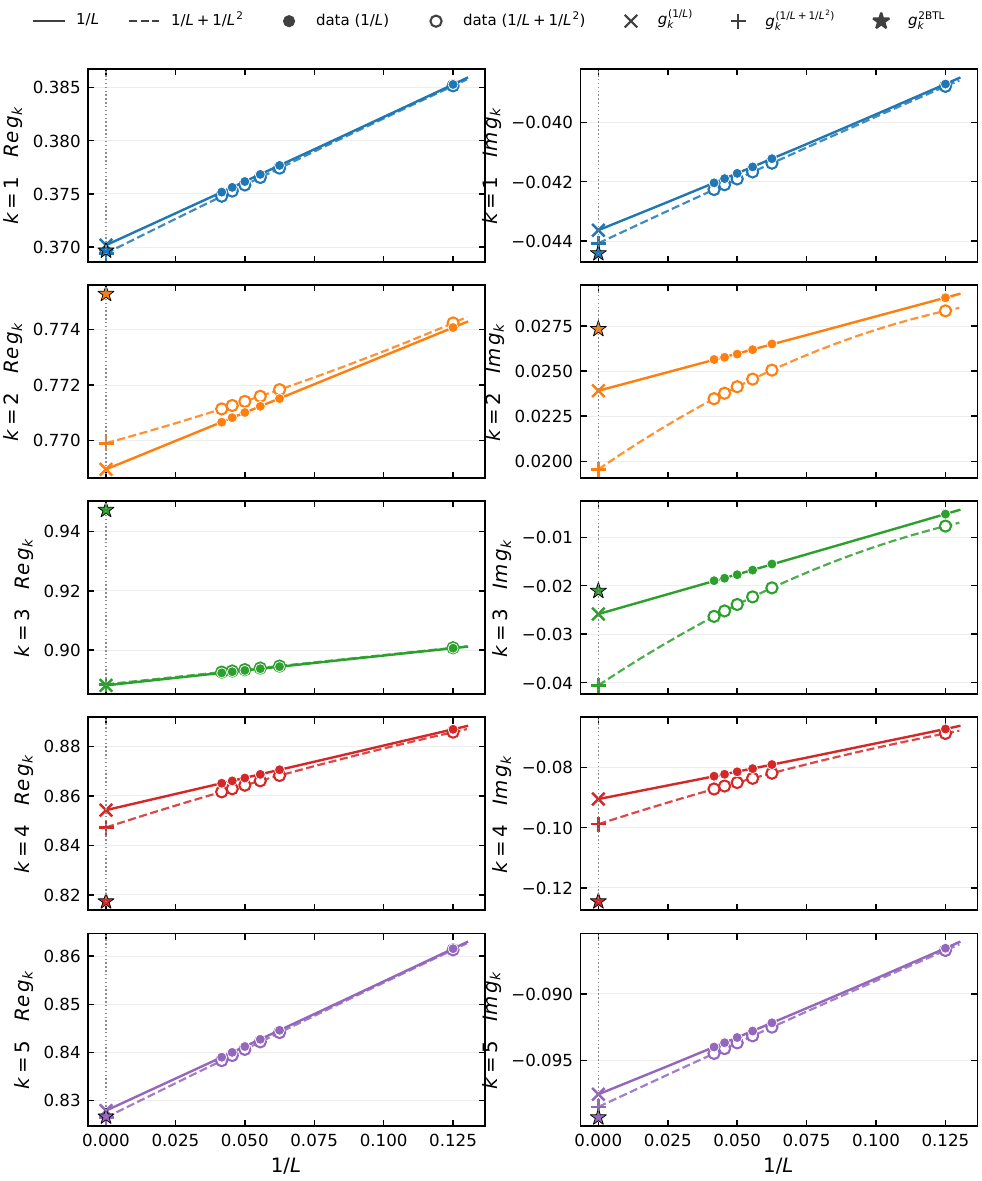}
  \caption{
  Ground-sector fit diagnostic for the complex Potts amplitudes
  $g_k=\sqrt{G_k}$.  The solid and dashed curves compare full complex fits of
  $-\log Z_k(L)$ with $1/L$ and $1/L+1/L^2$ corrections, respectively.  The data points are
  regularized by subtracting the corresponding fitted extensive term. 
  The filled and open circles correspond to
  the two fit ansatzes.  
  The crosses and plus signs mark the thermodynamic intercepts, while the stars denote the
  analytically continued 2BTL values.
  }
  \label{fig:supp-potts-ground-g-fit}
\end{figure}

The first excitation of the five-state Potts chain is fourfold degenerate in the Potts charge multiplet.  To
avoid choosing an arbitrary basis within this degenerate space, we define the projected amplitude to the
subspace rather than to a single excited state.  The four excited-state vectors used below are obtained with
the MPS quasiparticle ansatz~\cite{ChepigaMila2017}.  Let
$\{\left|\chi_{a,R}\right\rangle\}_{a=1}^{4}$ denote these states and define their Gram matrix
\begin{align}
  S_{ab}
  =
  \langle \chi_{a,L}|\chi_{b,R}\rangle .
\end{align}
The preparation-resolved projected amplitude to the first excited subspace is
\begin{align}
  Z_{k,{\rm ex}}(L)
  =
  \sum_{a,b=1}^{4}
  \langle \widetilde\Phi_k|\chi_{a,R}\rangle
  \left(S^{-1}\right)_{ab}
  \langle \chi_{b,L}|\Phi_k\rangle .
  \label{eq:supp-potts-subspace-proj}
\end{align}
We then form the normalization-independent ratio
\begin{align}
  R_{k,{\rm ex}}(L)
  \equiv
  \frac{Z_{k,{\rm ex}}(L)}{Z_k(L)} .
  \label{eq:supp-potts-ex-ratio}
\end{align}
The Gram matrices are well conditioned in the present data set: the largest condition number over
$L=8,16,18,20,22,24$ is about $6.5$.  Thus the subspace projection is numerically stable and does not rely
on resolving a particular charge basis inside the four-dimensional multiplet.

Figure~\ref{fig:supp-potts-first-ex-ratio} shows the complex ratio
$R_{k,{\rm ex}}(L)$ and compares the linear fits
\begin{align}
  R_{k,{\rm ex}}(L)
  =
  R_{k,{\rm ex}}{(\infty)}+\frac{A_{k,{\rm ex}}}{L}.
  \label{eq:supp-potts-ex-linear-fit}
\end{align}
with the two-correction fits obtained by adding $B_{k,{\rm ex}}/L^2$.
The extrapolated values are summarized in
Table~\ref{tab:supp-potts-ex-ratio}.
The finite-size drift is smooth on the accessible sizes, but the excited-sector ratios make the interpretation
of the intermediate preparations more restrictive than the ground-sector $g$ values alone.

\begin{table}[tb]
  \caption{
  \label{tab:supp-potts-ex-ratio}
  First-excited-subspace ratios for the Potts preparation family.  The numerical columns give the
  thermodynamic intercepts from the $1/L$ and $1/L+1/L^2$ fits in Fig.~\ref{fig:supp-potts-first-ex-ratio},
  with the parentheses denoting the propagated fitting uncertainty.  
  The last column gives the analytically
  continued 2BTL value.
  }
  \centering
  \begin{ruledtabular}
  \begin{tabular}{ccccc}
    $k$ &
    boundary &
    $R_{k,{\rm ex}}^{(1/L)}$ &
    $R_{k,{\rm ex}}^{(1/L+1/L^2)}$ &
    $R_{k,{\rm ex}}^{\rm 2BTL}$ \\ \hline
    $1$ &
    fixed &
    $8.206(2)+1.232(2)\,i$ &
    $8.232(5)+1.218(5)\,i$ &
    $8.2345+1.2850\,i$ \\
    $2$ &
    2-mixed &
    $1.842(6)+0.025(6)\,i$ &
    $1.863(4)+0.106(4)\,i$ &
    $1.4464+0.1256\,i$ \\
    $3$ &
    3-mixed &
    $0.263(4)-0.272(4)\,i$ &
    $0.207(2)-0.297(2)\,i$ &
    $0.7743-0.7141\,i$ \\
    $4$ &
    4-mixed &
    $0.013(1)-0.075(1)\,i$ &
    $-0.0046(5)-0.0662(5)\,i$ &
    $-0.3877+0.0356\,i$ \\
    $5$ &
    free &
    $\mathcal O(10^{-10})$ &
    $\mathcal O(10^{-10})$ &
    $0$
  \end{tabular}
  \end{ruledtabular}
\end{table}

\begin{figure}[tb]
  \centering
  \includegraphics[width=0.92\textwidth]{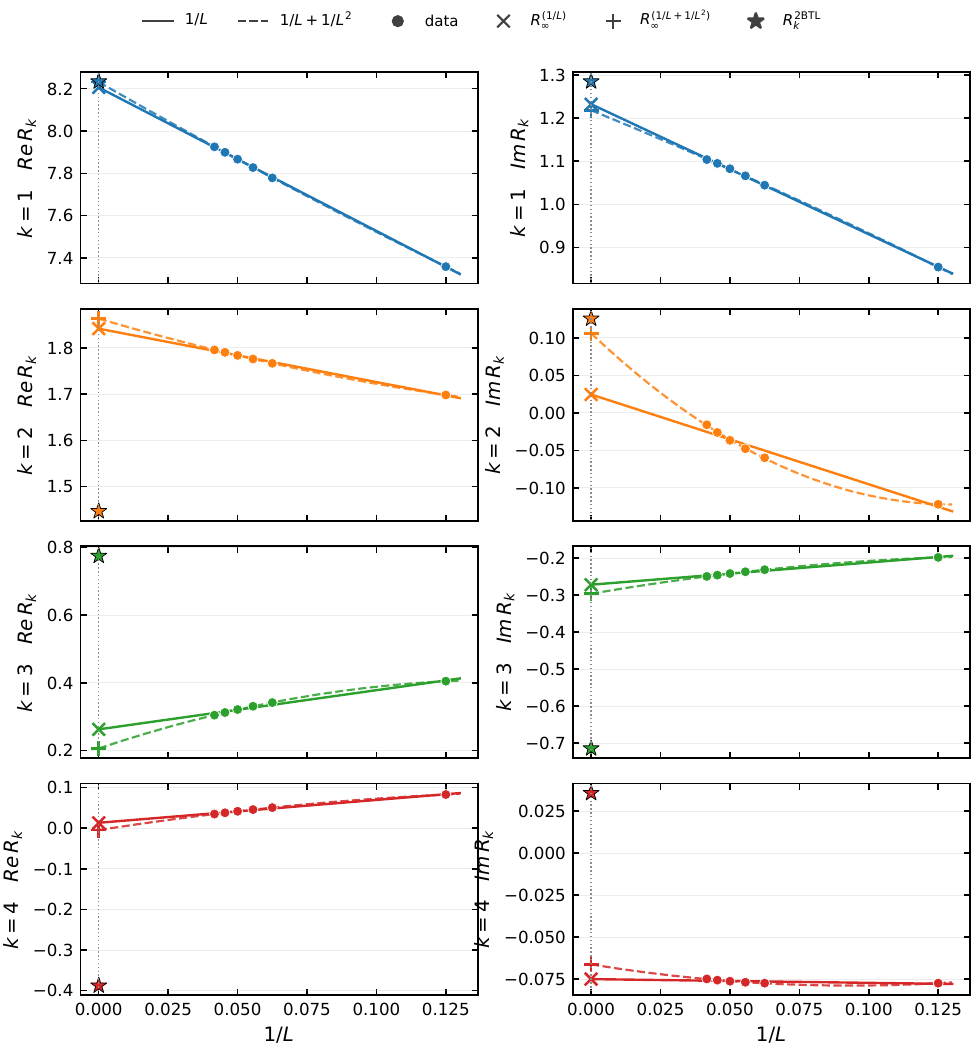}
  \caption{
  Fit comparison for the ratio
  $R_{k,{\rm ex}}(L)=Z_{k,{\rm ex}}(L)/Z_k(L)$, where
  $Z_{k,{\rm ex}}$ is the projection to the four-dimensional first-excited subspace.  
  The solid and dashed
  curves show the $1/L$ and $1/L+1/L^2$ fits, respectively.  
  The filled circles denote the data.
  The crosses and plus signs mark the corresponding extrapolated intercepts, 
  while the stars denote the analytically
  continued 2BTL values.  
  The nearly vanishing free-sector signal is omitted from this diagnostic plot.
  }
  \label{fig:supp-potts-first-ex-ratio}
\end{figure}

Combining the ground- and excited-sector diagnostics, the fixed and free preparations give the cleanest
agreement with the analytically continued 2BTL prediction; in particular, the ground-sector amplitudes satisfy
the Kramers--Wannier relation with high precision, and the free first-excited-subspace signal is consistent
with the vanishing 2BTL value.  By contrast, the remaining $k=2,3,4$ preparations show sector-dependent
discrepancies: $k=2$ is close at the level of the ground-sector amplitude, but its first-excited-subspace ratio
is incompatible with the 2BTL value, while $k=3,4$ already show visible deviations in the ground sector.  This is consistent with the conclusion of Ref.~\cite{TangBoundary2025}: the fixed
and free boundary conditions are robustly captured by the analytically continued 2BTL description, while the
intermediate blob-boundary continuations require additional care and do not directly describe the full
five-state Potts cylinder partition functions.

\end{widetext}

\end{document}